\documentclass[utf8]{FrontiersinHarvard} 
\usepackage{url,hyperref,microtype,subcaption}
\usepackage[onehalfspacing]{setspace}

\usepackage{longtable, tabularx}

\def\keyFont{\fontsize{8}{11}\helveticabold }
\def\firstAuthorLast{Chand {et~al.}} 
\def\Authors{Tarak Chand\,$^{1,2,*}$, Saurabh Sharma\,$^{1,*}$, Koshvendra Singh\,$^{3}$, Jeewan Pandey\,$^{1}$, Aayushi Verma\,$^{1,2}$, Harmeen Kaur\,$^{1}$, Mamta\,$^{1,4}$, Manojit Chakraborty\,$^{1,4}$, Devendra K. Ojha\,$^{3}$ and Ajay Kumar Singh\,$^{5}$
}

\def\Address{$^{1}$Aryabhatta Research Institute of Observational Sciences (ARIES) Manora Peak, Nainital-263001, India.\\
$^{2}$M.J.P. Rohilkhand University Bareilly-243006, India.\\
$^{3}$Department of Astronomy and Astrophysics Tata Institute of Fundamental Research (TIFR) Mumbai-400005, India.\\
$^{4}$Indian Institute of Technology (IIT) Roorkee, Roorkee-247667, India.\\
$^{5}$Department of Applied Physics/Physics Bareilly College, Bareilly-243006, India.}

\def\corrAuthor{Tarak Chand}

\def\corrEmail{tarakchands@gmail.com, tarakchand@aries.res.in}

\begin{document}
\onecolumn
\firstpage{1}

\title[Be 65]{Long-term investigation of an open cluster Berkeley 65} 

\author[\firstAuthorLast ]{\Authors} 
\address{} 
\correspondance{} 

\extraAuth{Saurabh Sharma \\ saurabh175@gmail.com, saurabh@aries.res.in}

\maketitle 

\begin{abstract}
We present a decade-long investigation of a poorly studied cluster, Berkeley 65 (Be 65), using deep optical data from the telescopes of ARIES, Nainital Observatory. We estimate its radius ($R_{cluster}$ = 1.6$^{'}$, aspect ratio of $\sim$1.1), distance (2.0 $\pm$ 0.1 kpc) and age ($\sim$160 Myrs). A clear turn-off point at $\sim$1.7 M$_\odot$ in the mass function suggests the escape of low-mass stars, and the lower photometric mass compared to the dynamical mass indicates ongoing disruption due to external forces.
Our long-baseline optical photometric data also identifies 64 periodic and 16 non-periodic stars in this region. We have presented the light curves and the classification of those variables. The periodic stars have periods ranging from  $\sim$0.05 days to $\sim$3.00 days and amplitude ranges from  $\sim$8 mmag to $\sim$700 mmag. The nonperiodic stars show variation from $\sim$30 mmag to $\sim$500 mmag. The periodic stars include main-sequence pulsating variables such as Slow Pulsating B-type, $\delta$ Scuti, RR Lyrae, and $\gamma$ Doradus.
We report a detached binary system and rotating variables similar to BY Draconis-type stars exhibiting variable brightness caused by starspots, chromospheric activity, and magnetic field-related phenomena.

\tiny
 \keyFont{ \section{Keywords:} galaxies: star clusters: general -- (ISM:) dust, extinction -- stars: variables: general -- (stars:) Hertzsprung-Russell and color-magnitude diagrams}.
\end{abstract}

\section{Introduction}

Most stars are believed to form in clusters within molecular clouds (\citealt{Lada_2003}) and are ideal sites to study star formation and stellar/galaxy evolution. The mass function distribution in a cluster (having a broad mass range) is an ideal statistical tool to investigate the star formation process. The dynamics of stars in the clusters, as well as the structure of clusters measured as a function of cluster age, hold important clues on the processes of galaxy evolution. The long-term monitoring of stars in the cluster also gives important clues on the internal physical processes related to their evolution, through their photometric variability behavior \citep{Lada_2003, Allen_2007prpl.conf..361A, Grasha_2017ApJ...842...25G, Grasha_2018MNRAS.481.1016G}. 

The star clusters, which are primarily formed in the Galactic disc, are subjected to disturbance, such as disc shock, spiral arm passage, molecular cloud encounters, etc. \citep{1958ApJ...127..544S,2012MNRAS.426.3008K}. As star clusters evolve, the removal of gas due to stellar feedback, along with gravitational interactions among stars and binary systems, gradually weakens the cluster's gravitational potential. This process results in the expansion of the cluster and can ultimately lead to its partial or complete dissolution into the surrounding galactic field \citep[e.g.,][]{1958ApJ...127..544S,2011MNRAS.415.3439D,2012MNRAS.427..637P,2013MNRAS.436.3727D,2013MNRAS.432..986P,2013A&A...559A..38P,2017A&A...600A..49B,2018MNRAS.473..849D}.
In the Galactic disc, open clusters typically have a lifespan of approximately 200 Myr before they begin to break apart \citep{2006A&A...446..121B,2013ApJ...762....3Y}. Clusters that survive beyond this period often develop distorted shapes and become less tightly bound, increasing their likelihood of breaking apart. After breaking apart, these former clusters give rise to moving groups and add stars to the field population \cite[see also,][]{Sharma_2020MNRAS.498.2309S}. Thus, open clusters are the sites to study the galaxy's evolution and dynamics \citep{2019ApJ...877...12T}.

Open clusters also harbor various types of variable stars, including pulsating, rotating, eclipsing binaries, and non-periodic variables,  in a broad mass spectrum.
Pulsating variables, such as  $\beta$ Cep, $\delta$ Scuti stars, $\gamma$ Doradus ($\gamma$ Dor) stars, and Slowly Pulsating B-type (SPB) stars, exhibit variable amplitudes ranging from a few millimagnitude (mmag) to hundreds of mmag over periods of hours to days. These stars are particularly valuable for investigating the internal structure and evolution as their variability arises from radial and non-radial pulsations \citep{Gautschy_1993,Kim_2001, Kang_2007,Chehlaeh_2018}.
On the other hand, rotating variables show periodic variability due to the modulation of stellar spots caused by stellar rotation. The variability amplitudes for these variables typically range from a few mmag to tens of mmag over days. The study of variability in these stars provides direct information about angular momentum evolution and magnetic activity \citep{Messina_2008}.
Overall, studying long-term photometric variability in stars gives us an essential insight into how different mechanisms evolve during the life span of stars. 

\begin{figure}[h!]
    \includegraphics[width=\textwidth]{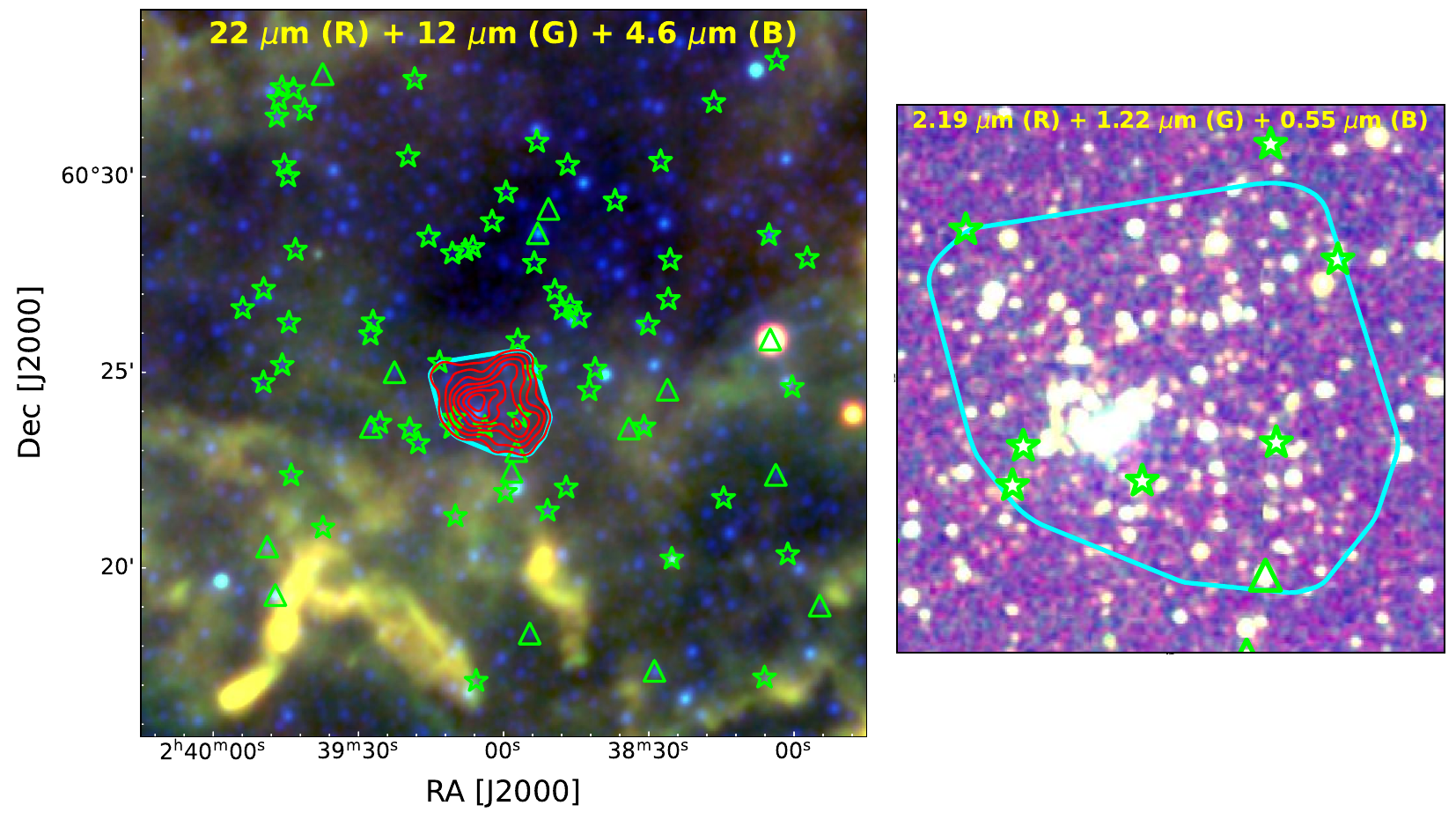}
    \caption{Left panel: Color composite image of the Be 65 cluster region covering $\sim$20$^{'}$ $\times$ 20$^{'}$ ~FoV, using the W2 (4.6 $\mu$m), W3 (12 $\mu$m), and W4 (22 $\mu$m) WISE images, shown as blue, green, and red colors, respectively. The red contours are the stellar iso-density contours generated using the nearest neighbor method from the 2MASS data (see Section~\ref{sec:cluster_structure}). The cyan color convex hull indicates the extent of the Be 65 cluster. The green open stars and triangles denote the identified periodic and non-periodic variables. Right panel: Zoomed-in image of the cluster region. This color-composite image comprises 2MASS $K$, $J$ band, and observed $V$ band images shown as red, green, and blue, respectively. }
    \label{fig:rgb}
\end{figure}

Thus, considering the above points, an intermediate-age open cluster (a few 100 Myr age, having most of the stars in the main sequence) located in the Galactic plane is ideally suited to study the dynamics and stellar evolution simultaneously.
Similar studies on star clusters have been carried out in the past decade. However, most of these studies are not always based on deep and long-term photometric data and usually lack membership determination based on high-quality proper motion (PM) data. This paper presents a photometric study of a poorly studied open cluster, Berkeley 65 (hereafter, Be 65; Figure~\ref{fig:rgb}). Be 65 is a Trumpler Class I 2p open cluster (\citealt{1966BAICz..17...33R}) in Cassiopeia constellation, located at $\alpha_{2000} = 02^{h}39^{m} 00^{s} $, $ \delta_{2000} = +60^{\circ} 25^{'} 00^{''} $ ($l = 135.8480^{\circ}$, b = $+0.2588^{\circ}$). 
The primary aim of the current study is to understand the dynamical state of this cluster precisely and to identify various types of variables in this region by using our decade-long photometric monitoring of this cluster along with high-quality data in archives, such as PM information from Gaia's third data release (\citealt{refId0}).

This paper is organized as follows: Section \ref{sec:obs} presents the observations and data reduction. Section \ref{sec:methodology} outlines the methods used to derive the cluster parameters and identify variable stars. In Section \ref{sec:res}, we discuss the physical and dynamical properties of the cluster Be 65, as well as the characteristics and classification of the identified variables. Finally, Section \ref{sec:summ} provides a summary of our findings and conclusions.

\section{Observations and Data Reduction}
\label{sec:obs}
\subsection{Optical photometric data}
The long-term photometric monitoring of Be 65 was done using the 1.3 m f/4 Devasthal Fast Optical Telescope (DFOT) at the Aryabhatta Research Institute of Observational Sciences (ARIES), located at Devasthal, India. The telescope has a 2K$\times$2K CCD camera with a field of view (FoV) of  $\sim$ $18^{'}.4\times18.^{'}4$. The CCD  has a $\sim$ 0.54 arcsec/pixel plate scale with a pixel size of $\sim$ 13.5 $\mu $m. The gain and read-out noise of the CCD are $\sim $2.0 electrons/ADU and $\sim$ 7.0 electrons, respectively \citep{Sagar_2012ASInC...4..173S}. About 1200 frames were taken in the Johnson $V$ filter over twenty-nine nights, from 2005 October 28 to 2022 November 04, including rigorous observation on two full nights and a few observations in the Johnson $U$, $B$, and Cousins $R, I$ bands. 
We have also observed the Be 65 cluster by 1.04 m ARIES Sampurnanand Telescope (ST), Nainital, in broadband $U$, $B$, $V$, $R_C$ and $I_C$ filters using the 2K $\times$ 2K CCD camera having a FoV of $13^{'}.5 \times13^{'}.5$ \citep{Sagar_2012ASInC...4..173S, Kaur_2020ApJ...896...29K}. The observing nights had good weather conditions, with a typical full width at half maxima (FWHM) for stellar images of around $1.5^{''}-2^{''}$.  
The detailed log of observations is given in Table~\ref{tab1}.

The basic data reduction, including image cleaning, photometry, and astrometry, was done using the standard procedure explained in \citet{Sharma_2020MNRAS.498.2309S} and \citet{Kaur_2020ApJ...896...29K}. 
The Be 65 field was observed on the same night (2005 November 04) as was NGC 6910 cluster \citep{Kaur_2020ApJ...896...29K} along with a standard field  \citep[SA98,][]{Landolt_1992AJ....104..340L}. The typical seeing value on that night was $\sim$ 7 pixels ($\sim$3$^{''}$), and observations were conducted under good atmospheric conditions. Thus, we have used the same calibration equations, as generated by \citet{Kaur_2020ApJ...896...29K}, to transform instrumental magnitudes of several bright stars in the Be 65 field into standard Vega systems.
For subsequent observations of Be 65 using the 1.3m telescope, it was not necessary to observe additional standard stars because the field already contained several bright stars with established standard magnitudes. Thus, to calibrate the 1.3m observations, a set of new transformation coefficients were directly derived from the standard and instrumental magnitudes of the local standard stars.
While there is a possibility that some of these stars may be variable, the statistical impact is minimized due to the large number of stars used. This approach also helps reduce errors associated with differences in airmass between standard stars and target stars, and it mitigates instrumental effects in the calibration equations.
The current observations can detect stars faint up to V$\sim$22 mag with photometric errors less than 0.1 mag. For the present analysis, we have used only those stars whose error in magnitude in different bands is less than 0.1 mag.
We transformed the pixel coordinates into celestial coordinates (RA and Dec) using the Graphical Astronomy and Image Analysis (GAIA\footnote{\url{https://astro.dur.ac.uk/~pdraper/gaia/gaia.html}}) tool with a root-mean-square (RMS) less than 0.5$^{''}$.


\subsection{Archival photometric data}

We have also used the photometric point source catalog from the following archives :
\begin{enumerate}
\item We obtained near-infrared (NIR) JHKs photometric data from the 2MASS All-Sky Point Source Catalog (\citealt{2003yCat.2246....0C}, \citealt{2006AJ....131.1163S}) for a stellar distribution study (see Section \ref{sec:metho_struc}). This catalog provides complete and reliable NIR photometry down to the 15.8, 15.1, and 14.3 magnitude limits in J (1.24 $\mu$m), H (1.66 $\mu$m), and $K_s$ (2.16 $\mu$m). We analyzed only sources with good photometric precision (photometric uncertainties less than 0.1 mag).

\item We have used data from Gaia's third data release, Gaia DR3 (\citealt{refId0}). Gaia DR3 provides astrometric positions, parallaxes, radial velocities, and proper motions for over 1.8 billion sources brighter than 21 mag in the G (0.33-1.05$\mu$m) photometric band. Precise magnitudes in three photometric bands G, $G_{BP}$(0.33-0.68$\mu$m), and $G_{RP}$(0.63-1.05 $\mu$m) are provided for up to 1.4 billion sources, along with astrophysical parameters and variability information for millions of objects. We have downloaded the Gaia DR3 data from the data archive\footnote{\url{https://www.cosmos.esa.int/web/gaia/data-release-3}}. We have used this data to discriminate between members and non-members of the clusters (see Appendix \ref{sec:memb}).
\end{enumerate}

\section{Methodology}
\label{sec:methodology}
\subsection{Structure of the cluster}
\label{sec:metho_struc}
As stars are less attenuated by dust and gas in NIR bands in comparison to optical bands, we have used the 2MASS All-Sky Point Source Catalog to investigate the distribution of stellar density in the Be 65 region, covering $\sim$20$^{'}$ $\times$ 20$^{'}$ FoV. A surface density map is produced using the nearest neighbor method as described in \cite{2005ApJ...632..397G, Gutermuth_2009ApJS..184...18G} and \cite{Sharma_2016}. Briefly, the local surface density was computed using a grid with a resolution of 5 pixels ($\sim$ 2.5 arcsec). At each grid point, the distance to the 20th nearest star was determined. Then the local surface density ($\sigma$) at each grid point [i,j] was estimated using the following equation:
\begin{equation}
    \sigma(i ,j) = \frac{N}{\pi \times (r_{N})^{2}}
\end{equation}
 where $r_{N}$ is the calculated distance of the nearest 20th stars from the grid point, and $N$ is the number of stars (i.e., 20) within a local area of radius $r_{N}$. Then the calculated surface density map was smoothed to a 3$\times$3 pixels$^2$ grid.
 
The resulting density contours, shown as red curves in the left panel of Figure~\ref{fig:rgb}, have the lowest contour level set at 1$\sigma$ above the mean stellar density ($8+2$ $stars/arcmin^2$) with contour intervals of 1$\sigma$ (2 stars/$arcmin^2$). The contours reveal a clear enhancement in stellar density in the central region.

\subsection{Extinction, distance, and age of the cluster}
\label{sec:extc_dis}

\begin{figure}[h!]
    \centering
    \includegraphics[width=8cm, height=8cm]{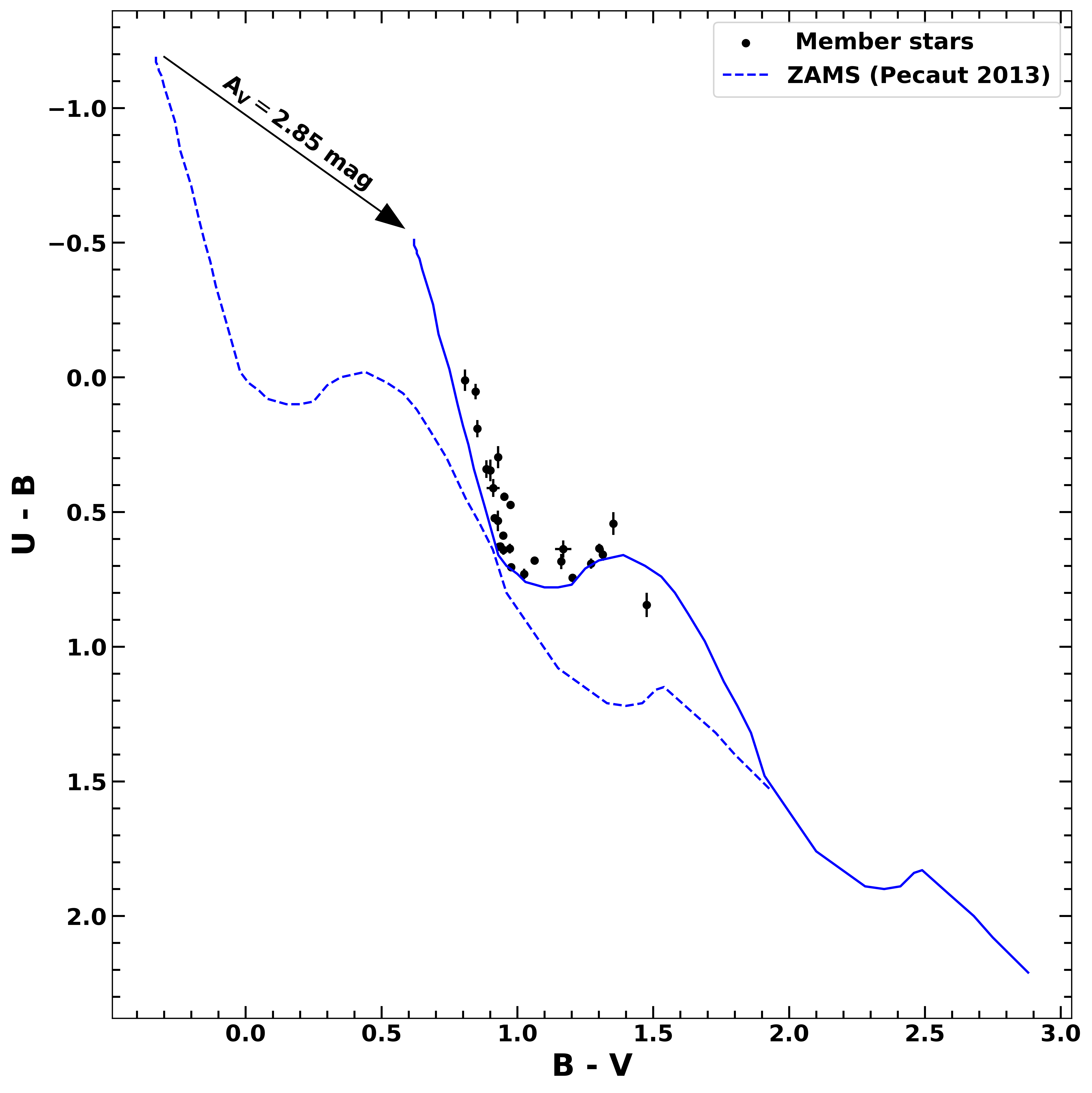}
    \includegraphics[width=8cm, height=8cm]{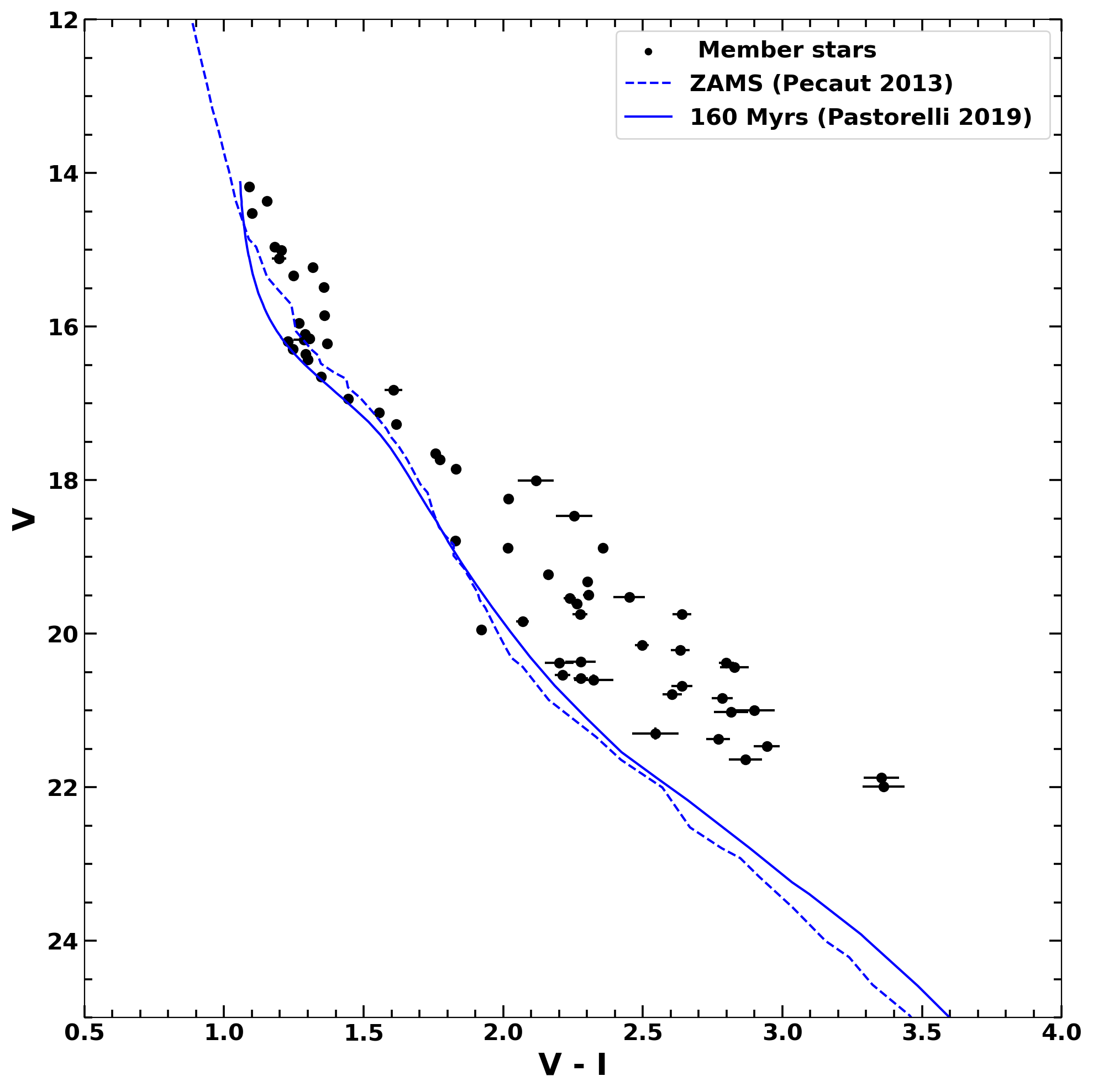}
    \caption{Left panel: $U-B$ V/s $B-V$ TCD for Be 65. The cluster member stars inside the convex hull are represented as black dots. The dashed blue line is the theoretical ZAMS curve taken from \citet{Pecaut_2013ApJS..208....9P}. The solid blue line is the reddened ZAMS curve along the reddening vector (black arrow) with $E(B-V) = 0.92$ mag. Right panel: $V$ vs. $(V-I)$ CMD for Be 65. The solid blue curve represents the theoretical isochrone taken from \citet{Pastorelli_2019} for age 160 Myrs and solar metallicity (i.e., Z=0.02), and the dashed blue curve is the theoretical ZAMS taken from \citet{Pecaut_2013ApJS..208....9P}. Both curves are corrected for distance and extinction value for Be 65 (i.e., 2.0 kpc and $ A_{V}=2.85$ mag).}
    \label{fig:ccd}
\end{figure}

We have used an optical $(U-B)$ versus $(B-V)$ two-color diagram (TCD) (see left panel in Figure~\ref{fig:ccd}) to estimate the reddening toward the Be 65 cluster \citep{Verma_2023, Sharma_2024AJ....167..106S}. We have plotted the probable member stars (see Appendix ~\ref{sec:memb}) that are within the convex hull as black dots 
along with zero-age main sequence (ZAMS; dashed line in blue color if Figure \ref{fig:ccd}) taken from  \citet{Pecaut_2013ApJS..208....9P}. We have shifted the ZAMS along the reddening vector (with a slope of $E(U-B)/E(B-V) = 0.72$, corresponding to $R_V$ = 3.1) to match the distribution of stars with minimum reddening in the cluster direction (blue solid curve in Figure \ref{fig:ccd}). Only stars of spectral type A or earlier have been selected for the reddening analysis to obtain a more reliable result. This choice is motivated by the desire to minimize several effects of factors like metallicity,  unresolved binaries, rotating stars, low mass pre-main sequence (PMS), and photometric errors, which can introduce uncertainties for later type stars (see \citealt{1994ApJS...90...31P} and ref therein). In this way, the minimum reddening value ($E(B-V)_{min}$) is estimated as 0.92 mag ($A_{V}$=2.85 mag). The approximate error in this estimate is 0.05 mag, as outlined in \cite{1994ApJS...90...31P}. 

To calculate the distance of the Be 65 cluster, we have selected 11 relatively brighter (V $<$16 mag) member stars (see Appendix~\ref{sec:memb}) which are inside the cluster extent, i.e., convex hull (see Section~\ref{sec:cluster_structure}). These selected member stars have a parallax error of less than 0.1 mas. The distances of these selected member stars were obtained from \cite{Bailer-Jones_2021AJ....161..147B}, and the mean of the distances of these stars has been considered as the distance of the cluster, i.e., 2.0 $\pm$ 0.1 kpc.
To further verify the estimated distance and extinction of the cluster, we applied the isochrone fitting method to its color-magnitude diagram (CMD), a technique that has been successfully utilized in numerous previous studies \citep{1994ApJS...90...31P, 2006A&A...449..151S, Sharma_2016, Pandey_2020ApJ...891...81P}. Specifically, we used the $V$ vs. ($V-I$) CMD, as shown in the right panel of Figure~\ref{fig:ccd}. We have shown the distribution of member stars (see Appendix~\ref{sec:memb}) within the convex hull as black dots. The blue dashed curve is the ZAMS curve taken from \cite{Pecaut_2013ApJS..208....9P} corrected for distance and minimum extinction.
A good match of ZAMS with the distribution of member stars provides additional confirmation of the cluster’s distance and extinction estimates.

The ages of young clusters are typically estimated using dereddened CMDs, either by comparing the most massive stars with post-main-sequence evolutionary tracks, if significant stellar evolution is evident, or by fitting PMS isochrones to the lower-mass, still-contracting population \citep{1994ApJS...90...31P, LATA_2014, Sharma_2020MNRAS.498.2309S, Rangwal_2023, Verma_2023}. The brighter blue stars typically serve as key constraints in isochrone fitting \citep{1994ApJS...90...31P}.
To determine the age of Be 65 cluster, we utilized the optical CMD (V versus V–I), presented in the right panel of Figure~\ref{fig:ccd} \citep{LATA_2014, Sharma_2020MNRAS.498.2309S, Rangwal_2023}. The ZAMS \citep{Pecaut_2013ApJS..208....9P}, corrected for the cluster’s distance and reddening, is overlaid as a dashed blue curve. The brightest star on the main sequence has an estimated mass of approximately 3.5 M$_\odot$. By visually fitting isochrones from \cite{Pastorelli_2019}, we find that an isochrone corresponding to an age 160 Myr (solid blue curve) aligns well with the observed member stars in the CMD. The uncertainty in the estimated age is influenced by factors such as differential reddening within the cluster and an extended period of star formation, both of which contribute to scatter in CMD \citep{1994ApJS...90...31P}. Given that we have corrected the isochrone using the minimum observed reddening value, and considering the variability in extinction across the cluster, we adopt a minimum age uncertainty of approximately 40 Myr.

\subsection{Mass function}
\label{sec:Mass_function}
The distribution of stellar masses during a star formation event is known as the initial mass function (IMF). The study of the IMF is crucial for understanding the star formation process and star clusters' subsequent chemical and dynamical evolution (\citealt{Kroupa_2002Sci...295...82K}). 
However, directly determining the IMF for a cluster is not possible due to the dynamic evolution of the stellar system. Therefore, we have estimated the present-day mass function (MF) of the Be 65 cluster. 
The MF is commonly represented by a power law equation, $\Phi \propto m^{\Gamma}$, and the slope of the MF is denoted as:

\begin{equation}
   \Gamma = \frac{d log(\Phi)}{d log (m)}\hspace{0.2cm}; \hspace{1.5cm}  \Phi = N (log (m))
    \label{eq:slop_mf}
\end{equation}

where $\Phi$ is the number of stars, corrected for completeness factor (CF), per unit logarithmic mass interval. 

\begin{figure}[h!]
    \centering
        \includegraphics[width=0.41\linewidth]{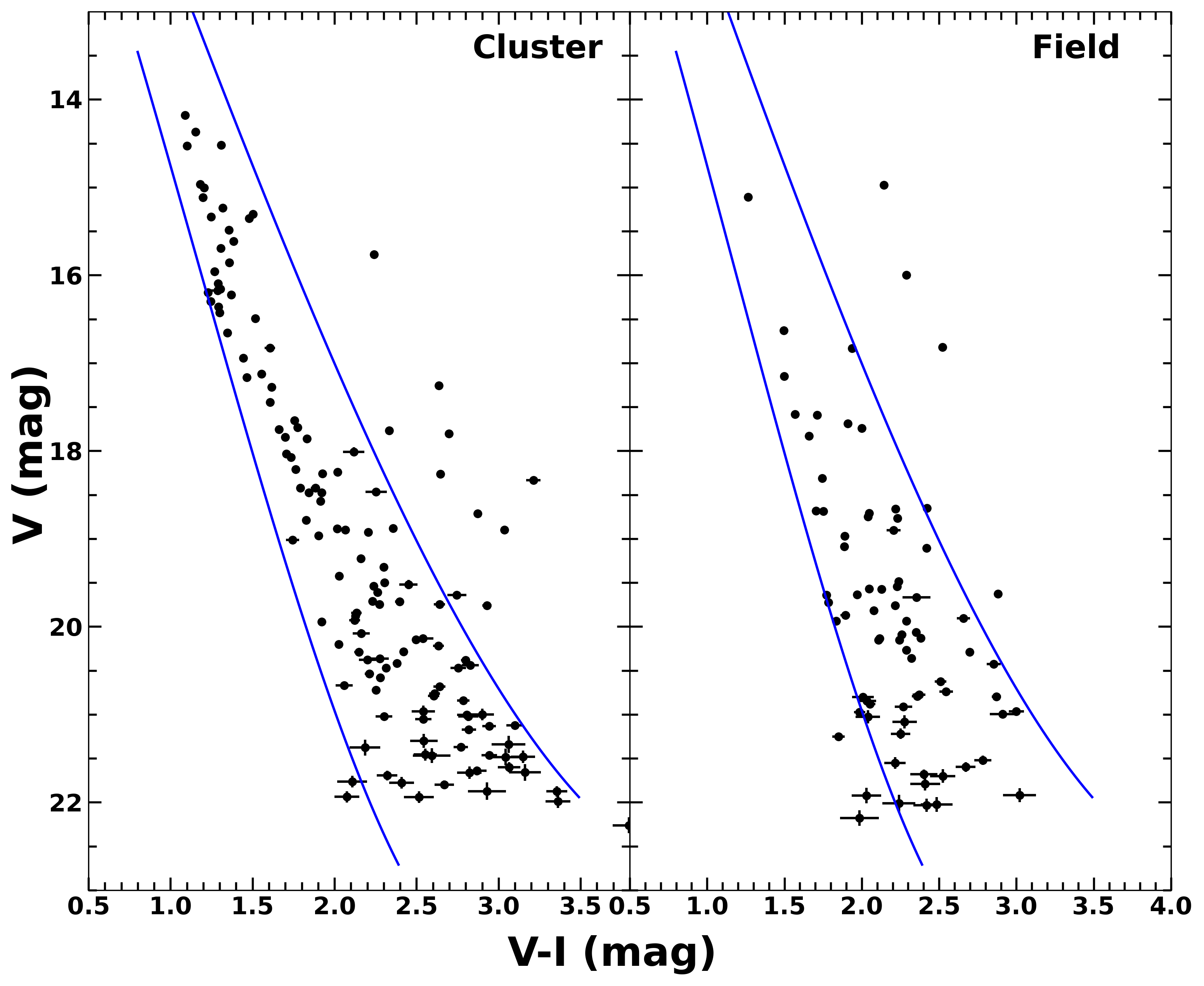}
        \includegraphics[width=0.50\linewidth]{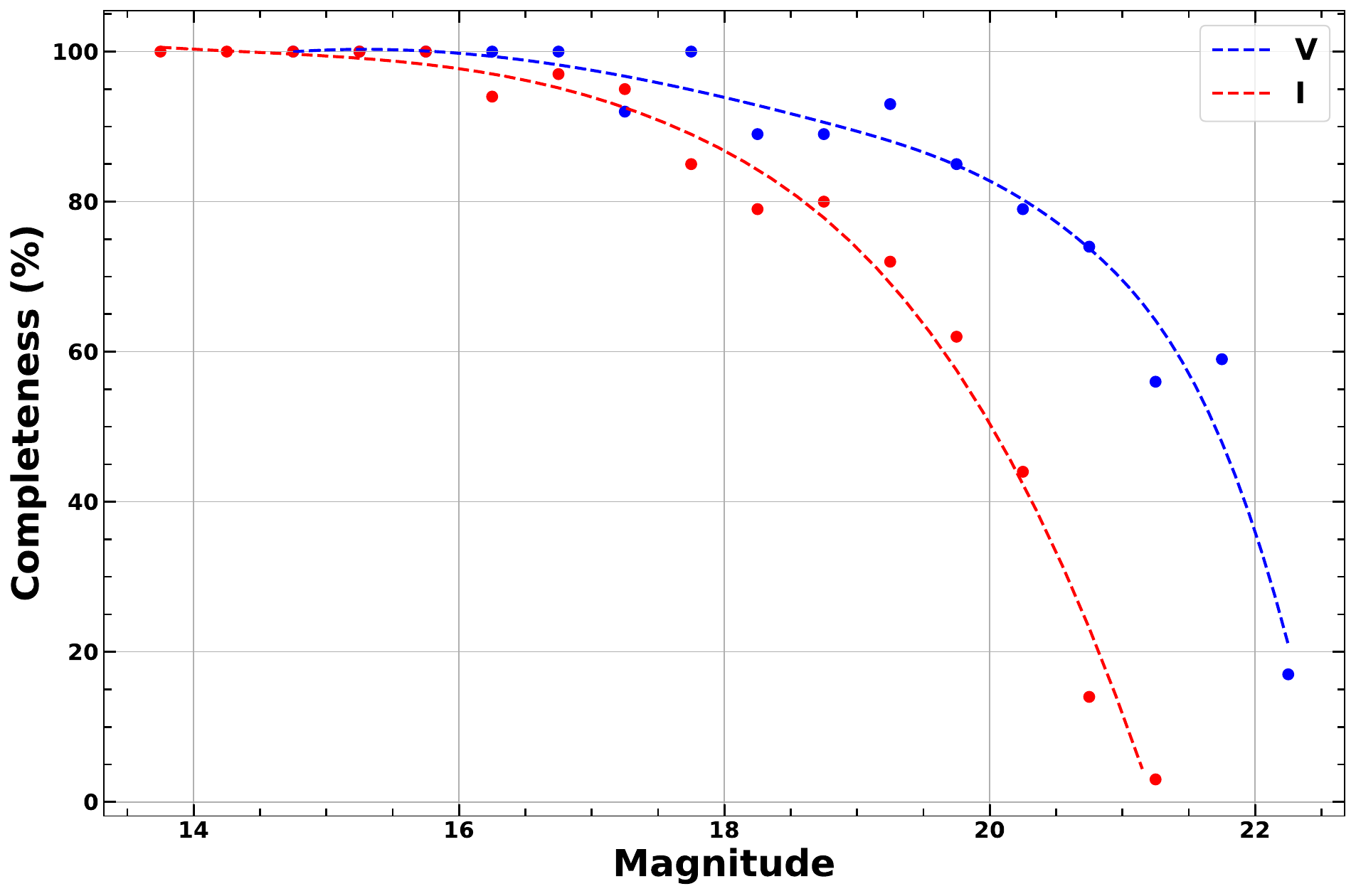}
    \caption{ Left panel: $V$ vs. ($V-I$) CMD for the stars in cluster and field regions. The blue line envelopes are created to select well-defined MS stars for LF and MF calculation for Be 65.
    Right panel:  Completeness factor for the cluster region of Be 65 as a function of magnitude. The red and blue dots are completeness factors derived using photometric $I$ and $V$ band data, respectively. The dashed lines are respective smoothed splines.}
    \label{fig:cft}
\end{figure}

To estimate the CF, we used the ADDSTAR routine from DAOPHOT II \citep{Stetson_1987PASP...99..191S, Stetson_1992ASPC...25..297S}. The detailed methodology is described in \cite{Sagar_1991A&A...250..324S}. In summary, artificial stars with known magnitudes and positions were randomly added to the original image. These modified frames were then re-reduced using the same reduction procedures as the original data. The CF as a function of magnitude was determined by calculating the ratio of recovered artificial stars to the total number of added stars within each magnitude bin. To better sample the fainter end of the luminosity function, a more number of artificial stars were added in the lower magnitude ranges. The number of added stars was limited to approximately 15$\%$ of the total star count in order to preserve the original crowding conditions.
The right panel of Figure~\ref{fig:cft} shows the CF versus magnitude plot in $V$ and $I$ bands for the cluster region. As expected, the CF decreases when we go towards the fainter end.

\begin{figure}[h!]
    \centering
    \includegraphics[width=0.5\linewidth]{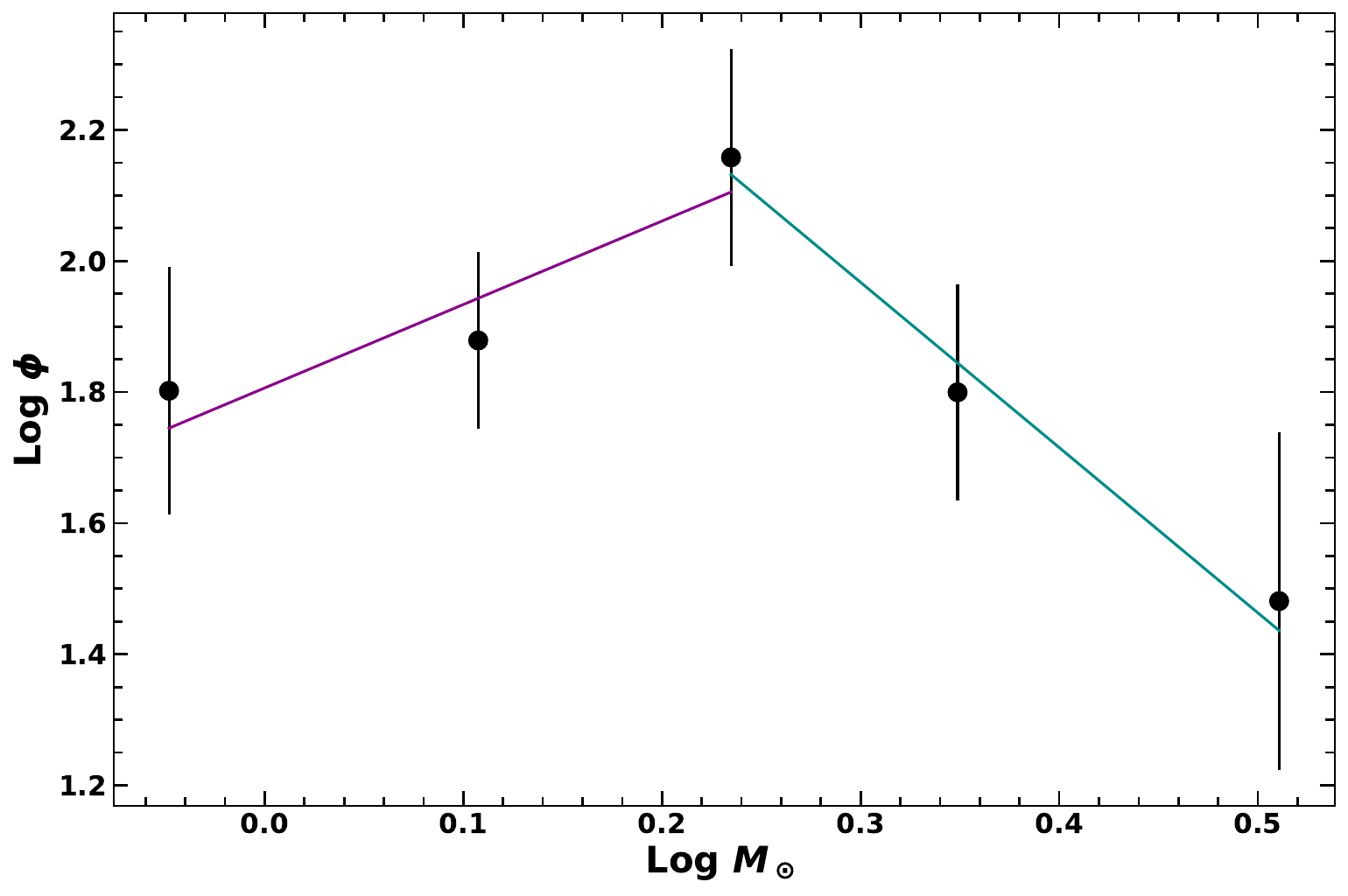}
    \caption{Mass function distribution for the Be 65 cluster. The error bars represented with MF data points (filled black dots) are $\pm \sqrt{N}$ errors. The solid cyan and magenta lines are the least-squares fits to the MF data points.}
    \label{fig:mass_f}
\end{figure}

The left panel of Figure~\ref{fig:cft} presents the CMDs for stars in both the cluster and field regions. We defined an envelope to select well-defined MS stars in each region (cf. \citealt{Sharma_2008}), to reduce field star contamination in the cluster region. To construct the cluster’s luminosity function (LF), we subtracted the CF-corrected star counts in the field region from those in the cluster region, in different mag bins within the envelopes in the CMDs. Finally, the magnitude bins in the LF were converted into mass bins using a 160 Myr isochrone from \cite{Pecaut_2013ApJS..208....9P}, and the resulting MF distribution is shown in Figure~\ref{fig:mass_f}.

\subsection {Identification  of variables}
\label{sec:identi_var}
\begin{figure}[h!]
    \centering
    \includegraphics[width=0.95\textwidth]{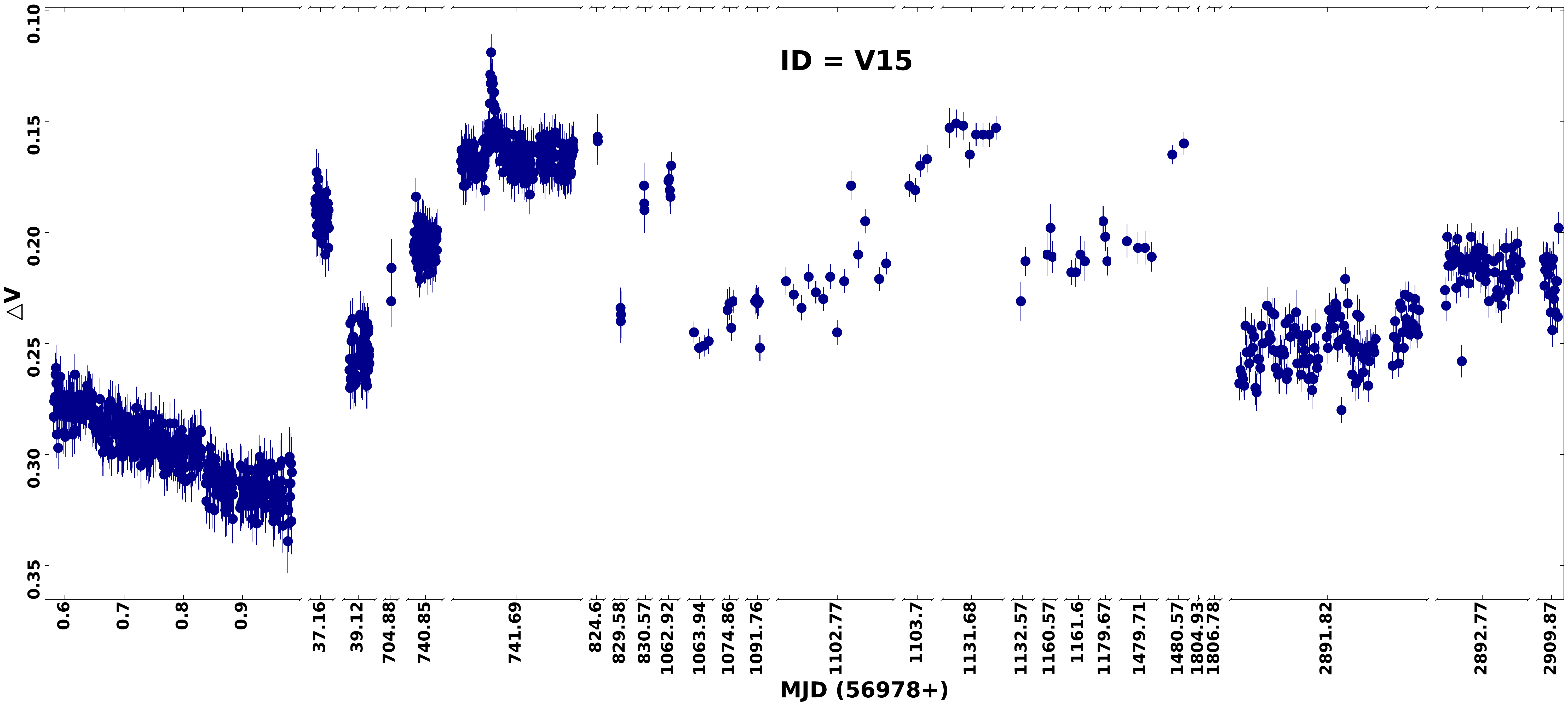}
    \includegraphics[width=0.95\textwidth]{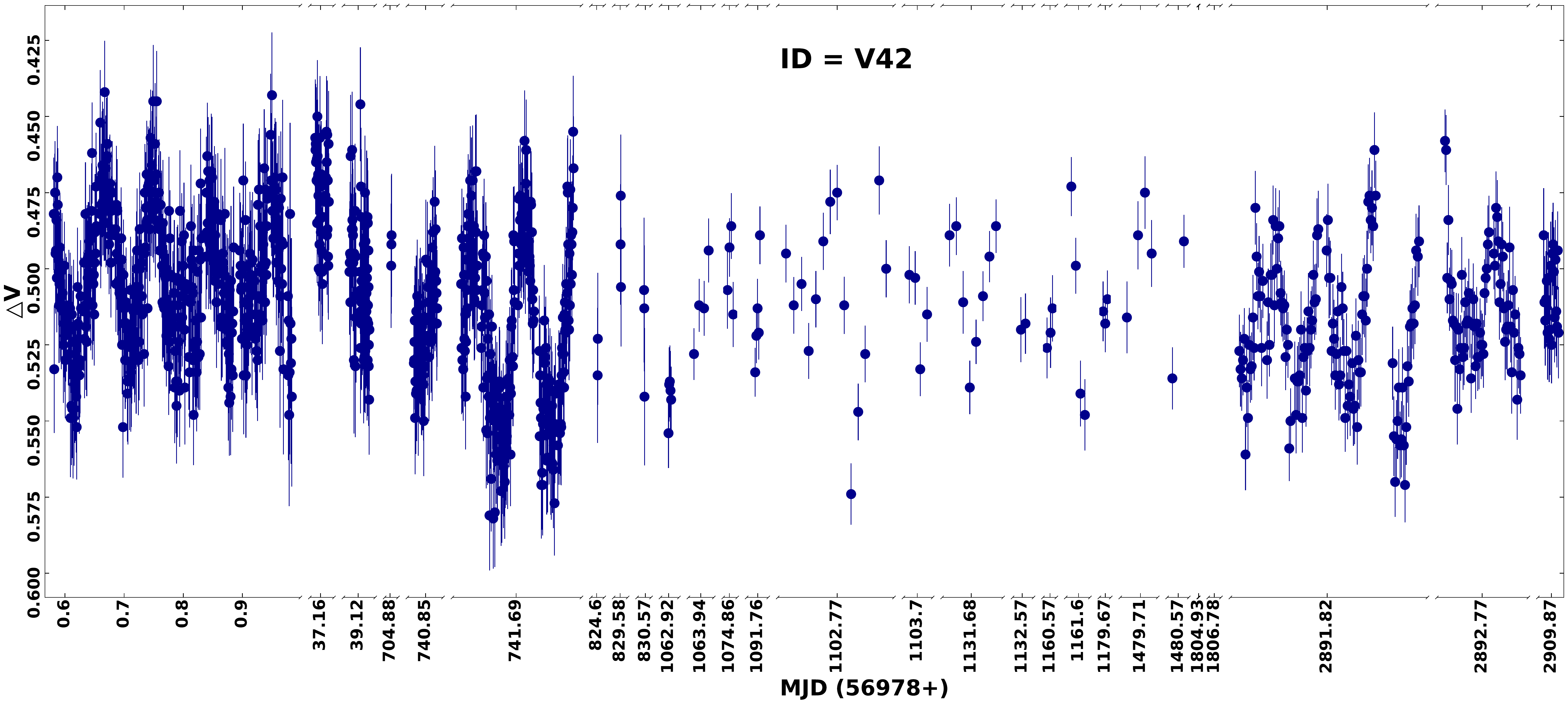}
    \caption{Samples of the light curves: in the upper panel, the light curve of a non-periodic variable (V15) is shown. The light curve of a periodic variable (V42) is shown in the lower panel.}
    \label{fig:lc}
\end{figure}

\begin{figure}[h!]
    \centering
    \includegraphics[width=0.5\textwidth]{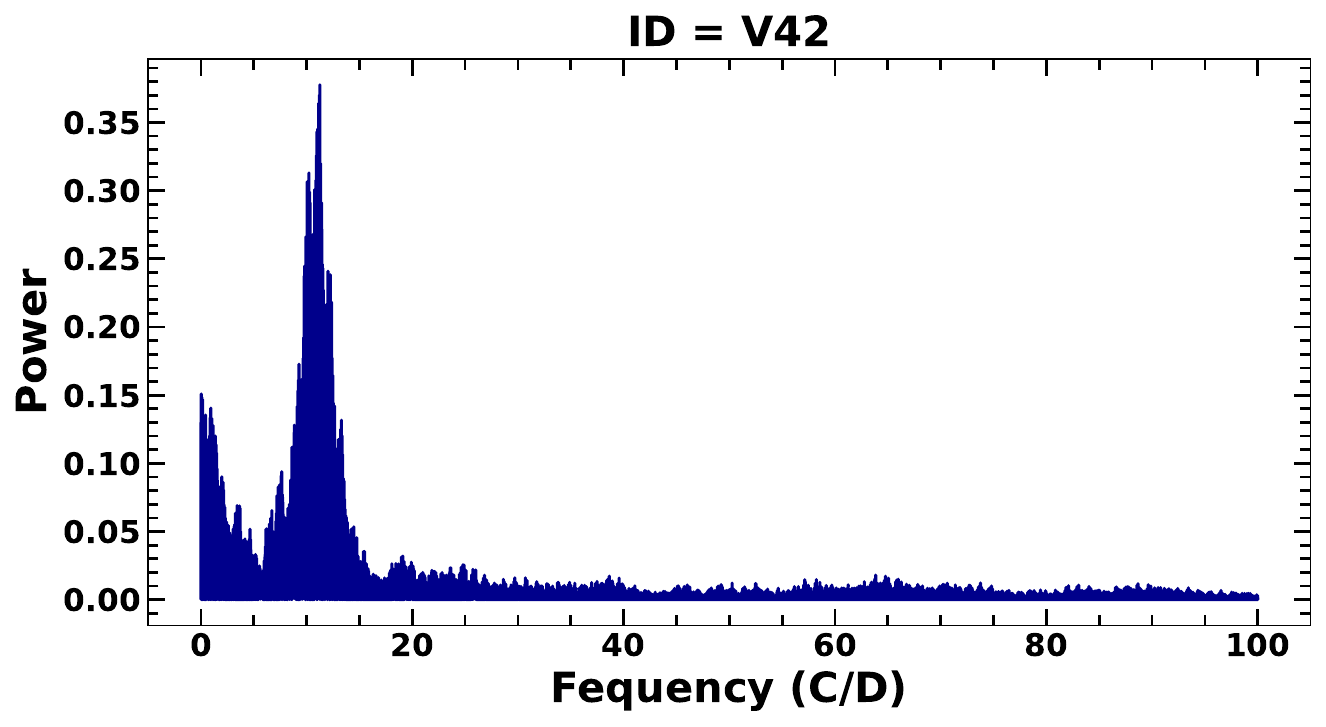}
    \includegraphics[width=0.5\textwidth]{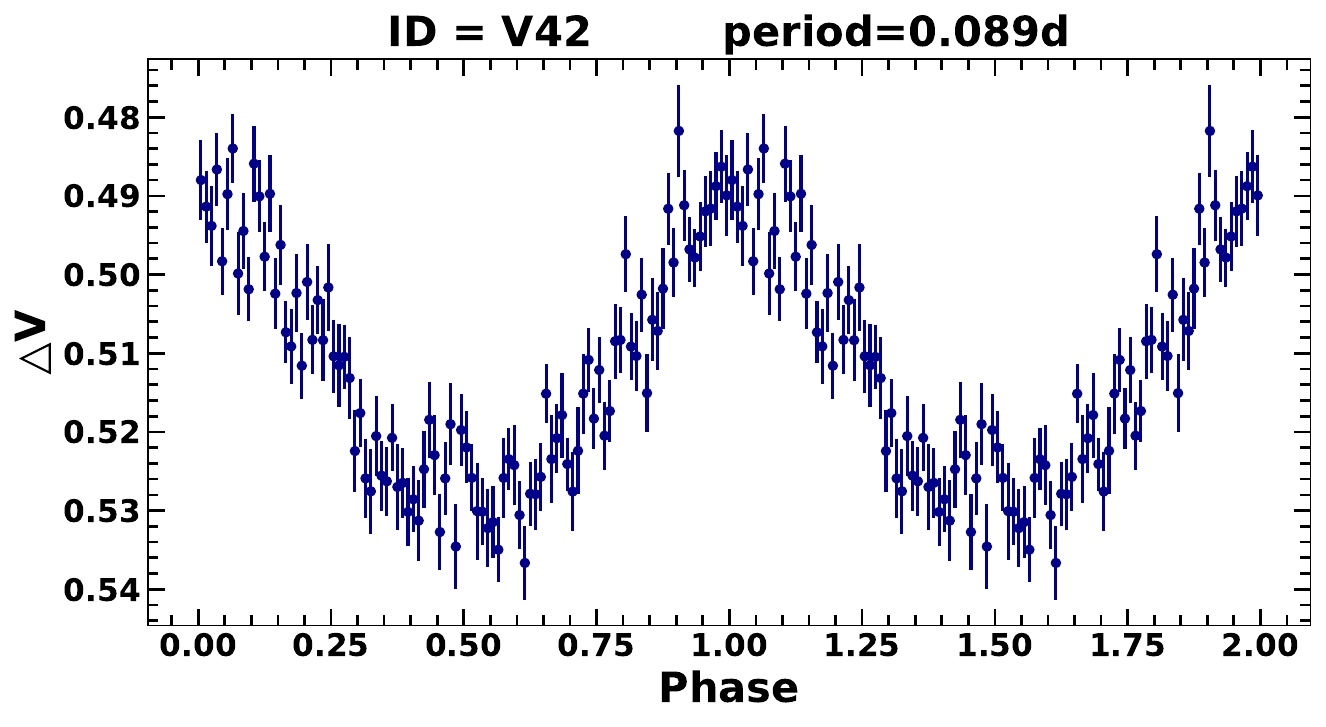}
    \includegraphics[width=0.5\textwidth]{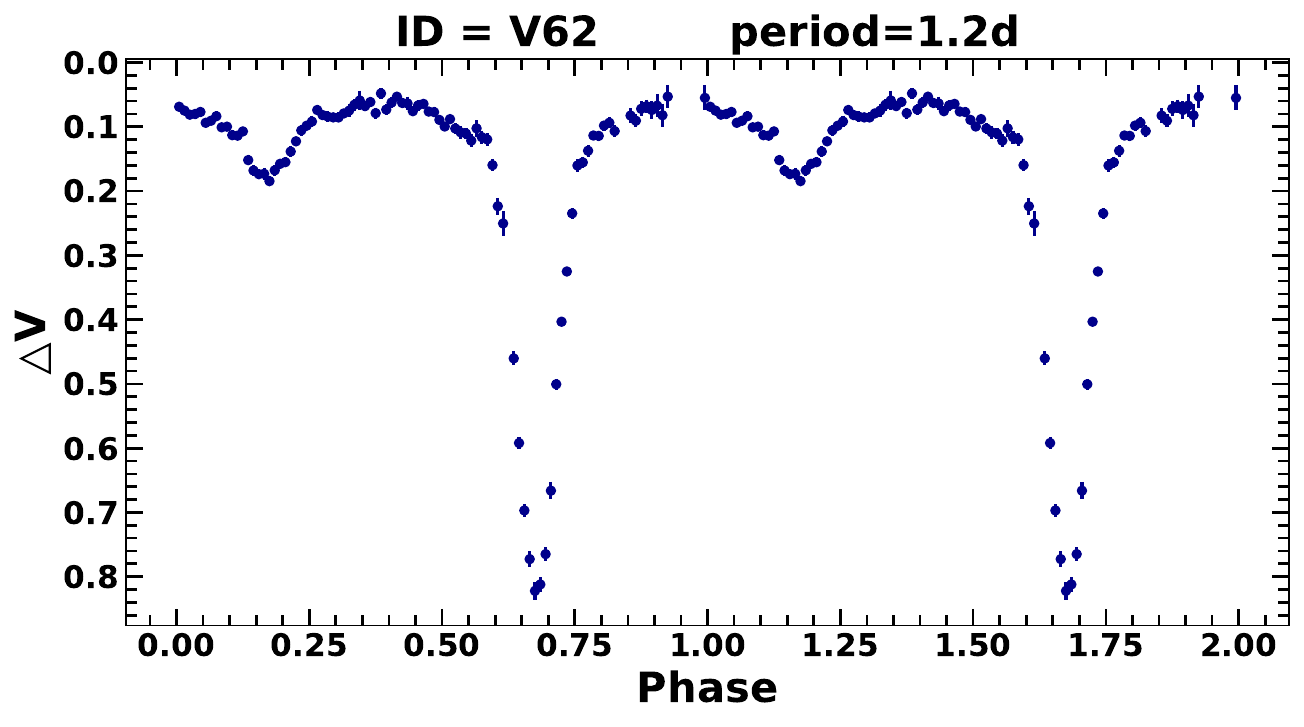}
    \caption{Upper panel: The power spectrum of a periodic star (V42) derived using Lomb-Scargle periodogram \citep{Lomb_1976Ap&SS..39..447L, Scargle_1982ApJ...263..835S}. Middle panel: the phase-folded light curve of the same periodic star (V42) using the period as 0.089 days. Lower panel: The phase-folded light curve of a detected eclipsing binary (V62); approx. 1.2 days is used as a period to fold the light curve.}
    \label{fig:phase}
\end{figure}

\begin{figure}[h!]
    \centering
    \includegraphics[width=0.45\textwidth]{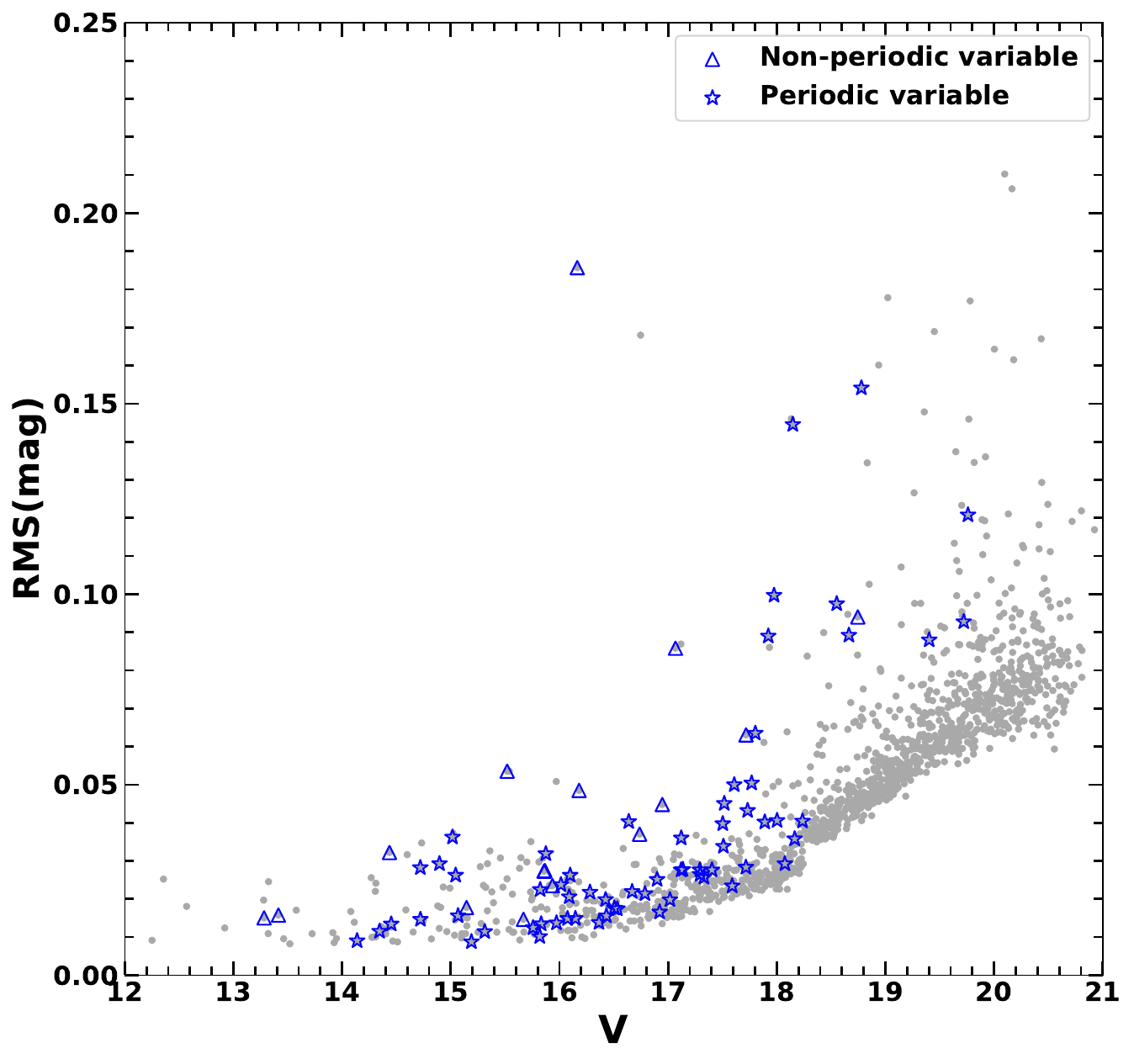}
    \caption{RMS dispersion of magnitudes for all the target stars
as a function of their V magnitude. The grey dots represent the stars in FoV towards the Be 65 cluster, while the blue open triangles and stars represent non-periodic and periodic variables towards the Be 65 cluster.}
    \label{fig:mag_rms}
\end{figure}

Our long-term photometric monitoring of Be 65 is ideal for identifying this region's long-period and short-term variables.
We have used differential photometric techniques to identify variable stars in the field. This technique is particularly useful for removing the star's brightness variation due to atmospheric conditions on the night of observation and instrumental signatures. It is important to note that the error in calculated magnitude increases as the stars become fainter.
Therefore, a brighter star cannot be used as a comparison star for the faint stars, as larger magnitude errors can obscure the star's variation. To overcome this, we divided all the stars into different magnitude bins with a bin size of 1 mag. We computed the magnitude differences for each magnitude bin for every possible pair of stars in a frame. We selected the comparison stars for which the standard deviation in the magnitude difference was the minimum. 
After selecting the comparison star, we determined the magnitude difference between the comparison star and target stars in each magnitude bin. Subsequently, we generated light curves (LCs) by plotting the resulting $\bigtriangleup$V (magnitude difference) against the Modified Julian Date (MJD). We visually inspected all the LCs in the initial step to identify any signs of variability. If a star displayed a consistent visual variation greater than the scatter observed in the comparison star, it was classified as a variable star. The Lomb-Scargle periodogram (\citealt{Lomb_1976Ap&SS..39..447L}, \citealt{Scargle_1982ApJ...263..835S}) is then used to determine the period of all periodic variable stars. As data is highly unevenly sampled, false periods can also show up in the periodogram. Therefore, we have confirmed the period of the periodic variable by plotting their phase-folded LCs. The phase values in these phase-folded LCs are binned so that each bin, with a bin size of 0.01, contains a single data point. In this way, we have identified 64 periodic stars toward the Be 65 cluster. 
For those variables for which no periods were estimated (16) successfully, we call them non-periodic variables. 
In Figure~\ref{fig:lc}, we show the sample LCs of periodic (V42) and non-periodic (V15) stars. In Figure~\ref{fig:phase}, the power spectrum and phase-folded LC (in the upper and middle panel, respectively) of a periodic star V42 is shown, while in the lower panel, the phase-folded LC of a detected eclipsing binary (for detail, see Section~\ref{sec:class_var}) is shown. The LCs of all variable stars are provided in supplementary data, 
and their derived parameters (e.g., coordinates, V magnitudes, periods, amplitudes, etc.)  are given in Table~\ref{tab:var}. Variables with distances (taken from \citealt{Bailer-Jones_2021AJ....161..147B}) falling within the one-sigma range of the estimated cluster distance (as discussed in Section~\ref{sec:extc_dis}) are classified as member stars (35) in Table~\ref{tab:var}, while the remaining are considered field stars (45). Although many of the member variables lie outside the apparent cluster extent (see Figure~\ref{fig:rgb}), this may be due to their dispersal from the cluster as a result of an ongoing disruption process.
Figure~\ref{fig:mag_rms} illustrates the RMS dispersion of magnitudes for all the observed stars as a function of their V magnitude. As expected, there is an increase in dispersion towards the fainter end of the magnitude range. The identified variables are represented by open blue stars (periodic) and open blue triangles (non-periodic). In this figure, some stars with a very high RMS are not classified as variables. This could be due to unusually high photometric errors (e.g., bad pixels within the star, the presence of a nearby brighter star, cosmic correction residuals).

\section{Results}
\label{sec:res}
\subsection{Physical properties of the Be 65 cluster}
\label{sec:cluster_structure}
Using the surface density estimates, the peak density is found to be located at $\alpha_{2000} = 02^h  39^m 05^s $, $ \delta_{2000} = +60^{\circ} 24^{'} 30^{''} $. As the cluster does not have perfectly spherical symmetric morphology, we define its area using a convex hull\footnote{Convex hull is an irregular polygon enclosing all points in a grouping with internal angles between two contiguous sides of less than 180 $\deg$.} , represented by the cyan-colored curve in the right panel of Figure~\ref{fig:rgb} \citep[see][for details]{2006A&A...449..151S}.
 The Qhull\footnote{Barber, C. B., D.P. Dobkin, and H.T. Huhdanpaa," The Quickhull Algorithm for Convex Hulls," ACM Transactions on Mathematical Software, 22(4):469-483, Dec 1996, \url{www.qhull.org}.} program is used to compute the convex hull for the stars located within the lowest density contour. We estimated the area of cluster $A_{cluster}$, as the area under the convex hull normalized by a geometrical factor that considers the ratio of objects contained within the convex hull versus those situated on its boundary \citep[cf.][and references therein]{Sharma_2020MNRAS.498.2309S}. Then, the cluster radius, $R_{cluster}$, is determined as the radius of a circle with the same area as $A_{cluster}$. Additionally, we have computed the circular radial size, $R_{circle}$, which is defined as half of the maximum distance between any two members within the cluster. It represents the radius of the smallest possible circle that contains the entire grouping. In this way, we have estimated the $A_{cluster}$, $R_{cluster}$ and aspect ratio $ R^2_{circle}/R^2_{cluster}$ \citep{Gutermuth_2009ApJS..184...18G} for the  Be 65 cluster as 8.3 $arcmin^2$, 1.6$^{'}$ and 1.1, respectively.

The minimum reddening value ($E(B-V)_{min}$) towards Be 65 cluster is estimated as $0.92\pm0.05$ mag ($A_{V}$=2.85 mag). The seemingly less scattered distribution of stars in the TCD hints toward the negligible differential reddening or gas/dust in this cluster. This hints towards the evolved phase of this cluster. Our PM and CMD fitting analysis confirms the distance of this cluster as 2.0 $\pm$ 0.1 kpc. The cluster's age, estimated from the turn-off point of the brightest member star for the cluster, is estimated as $\sim160$ Myrs.

Although there is a scatter in the MF distribution, a distinct change in slope can be observed at log$(M_\odot) \approx 0.23$ (or $ M \approx 1.7 M_\odot)$. This break in the MF slope has also been observed in previous studies of other clusters  \citep[e.g.,][]{ Sharma_200710.1111/j.1365-2966.2007.12156.x, Jose_2008}. 
We calculated distinct $\Gamma$ on either side of the observed break using a least-squares fitting method. The MF slope in the relatively higher-mass ($1.7 M_\odot \leq M \leq 3.5 M_\odot $) regime is found to be -2.52 $\pm$ 0.15, whereas the MF slope in the low-mass regime ($ 0.9M_\odot \leq M \leq 1.7 M_\odot$) is estimated as 1.27 $\pm$ 0.33.

\subsection{Dynamical state of the cluster}
\label{sec:dynamics}
In our study, we used the minimum spanning tree (MST\footnote{The MST is a network of branches that connects a set of points while minimizing the total branch length and avoiding any loops  \citep{Battinelli_1991A&A...244...69B}. This algorithm has lately become a popular tool to search for clusters of stars since it is independent from the star’s density number \citep{Gutermuth_2009ApJS..184...18G, Chavarr_2014MNRAS.439.3719C, Sharma_2016}.}) method \citep{Allison_2009MNRAS.395.1449A, Olczak_2011A&A...532A.119O, Dib_2018MNRAS.473..849D}, to investigate mass segregation within the Be 65 cluster region. We used the mass segregation ratio (MSR, $\Gamma_{MSR}$) to quantify mass segregation. The $\Gamma_{MSR}$ was calculated by constructing MSTs for two groups: the massive member stars (with magnitudes G$<$17) and an equal number of randomly selected stars from all cluster members within the cluster region. For each group, the total edge length of the MST was computed over 100 iterations, and the mean edge lengths were determined. The $\Gamma_{MSR}$ was then obtained by taking the ratio of the mean edge length of the random sample to that of the massive stars. The uncertainty in the $\Gamma_{MSR}$ was estimated as the standard deviation of the MST edge lengths of random samples \citep[for details, see][]{Sharma_2020MNRAS.498.2309S}. 
The calculated value of $\Gamma_{MSR}$ for the Be 65 cluster is $1.1\pm1.2$. This indicates the presence of mass segregation, suggesting a concentration of massive stars towards the central region of the cluster \citep[for further details, see][]{Sharma_2020MNRAS.498.2309S, Sharma_2023JApA...44...46S, Kaur_2023JApA...44...66K}. To explore whether the observed mass segregation in Be 65 originates from the dynamical relaxation or is inherent in its formation, we calculated the dynamical relaxation time, $T_E$, as $\sim$4.4 Myr based on the sample of member stars in the Be 65 cluster \citep[for more information, refer to][]{Sharma_2020MNRAS.498.2309S}. The probable age of the cluster is calculated as 160 Myr (see Section~\ref{sec:extc_dis}), which is much higher than the $T_E$. Thus, the cluster can be considered dynamically relaxed. Even considering a loss of 50$\%$ of stars due to data incompleteness, the dynamical relaxation time will be $T_E \sim 7.5$ Myr. However, this remains considerably smaller than the age of the Be 65 cluster (i.e., $\sim$ 160 Myr). This suggests that the observed mass segregation in this cluster could result from the dynamical evolution of stars. 

One of the consequences of the mass segregation process is the increased vulnerability of the lowest-mass members to be ejected from the system \citep[e.g., see][]{Mathieu_1984ApJ...284..643M}. 
Thus, we estimated the tidal radius `$r_t$' of Be 65, following the methodology outlined by \citet{Sharma_2020MNRAS.498.2309S}. To calculate the tidal radius, we first derived the cluster stars' MF distribution using the member stars' LF (as described in Section~\ref{sec:Mass_function}). The total mass of the Be 65 cluster is estimated to be 82 $M_{\odot}$ up to the completeness limit of the observed photometric V band data, corresponding to 0.9 $M_{\odot}$. Based on these calculations, the tidal radius is approximately 6.3 pc. If we missed 50$\%$ of the cluster mass in the lower mass bins due to data incompleteness, the resulting tidal radius `$r_t$' would be approximately 8 pc. 
We then compared the photometric mass with the dynamical mass (M$_{dyn,tid}$) for stars within the tidal radius to quantify this cluster's dynamical status further. The dynamical mass is calculated by,

\begin{equation}
{M_{dyn}} \sim {\frac{r_t\sigma^2_{3D}}{G}}   
\end{equation}

where $r_t$ is the 3D tidal radius, and $\sigma_{3D}$ is the
3D velocity dispersion \citep{2006MNRAS.369.1392F,2019ApJ...877...12T}.
Assuming an isotropic velocity distribution within the tidal radius, $\sigma^2_{3D}$ is 3 times the 1D velocity dispersion $\sigma^2$ of cluster members. Using the typical radial velocity dispersion of 1 km s$^{-1}$ for open clusters \citep{1989AJ.....98..227G}, the M$_{dyn,tid}$ for Be 65 cluster comes out to be  $\sim$5581 M$_\odot$. This is much higher than the estimated photometric mass of Be 65 within the tidal radius, i.e., $\sim$164 M$_\odot$, suggesting that this cluster has lost stellar mass and, thus, is under the process of disruption.

\subsection{Physical properties of the variables}
\begin{figure}[h!]
    \centering
    \includegraphics[width=0.45\linewidth]{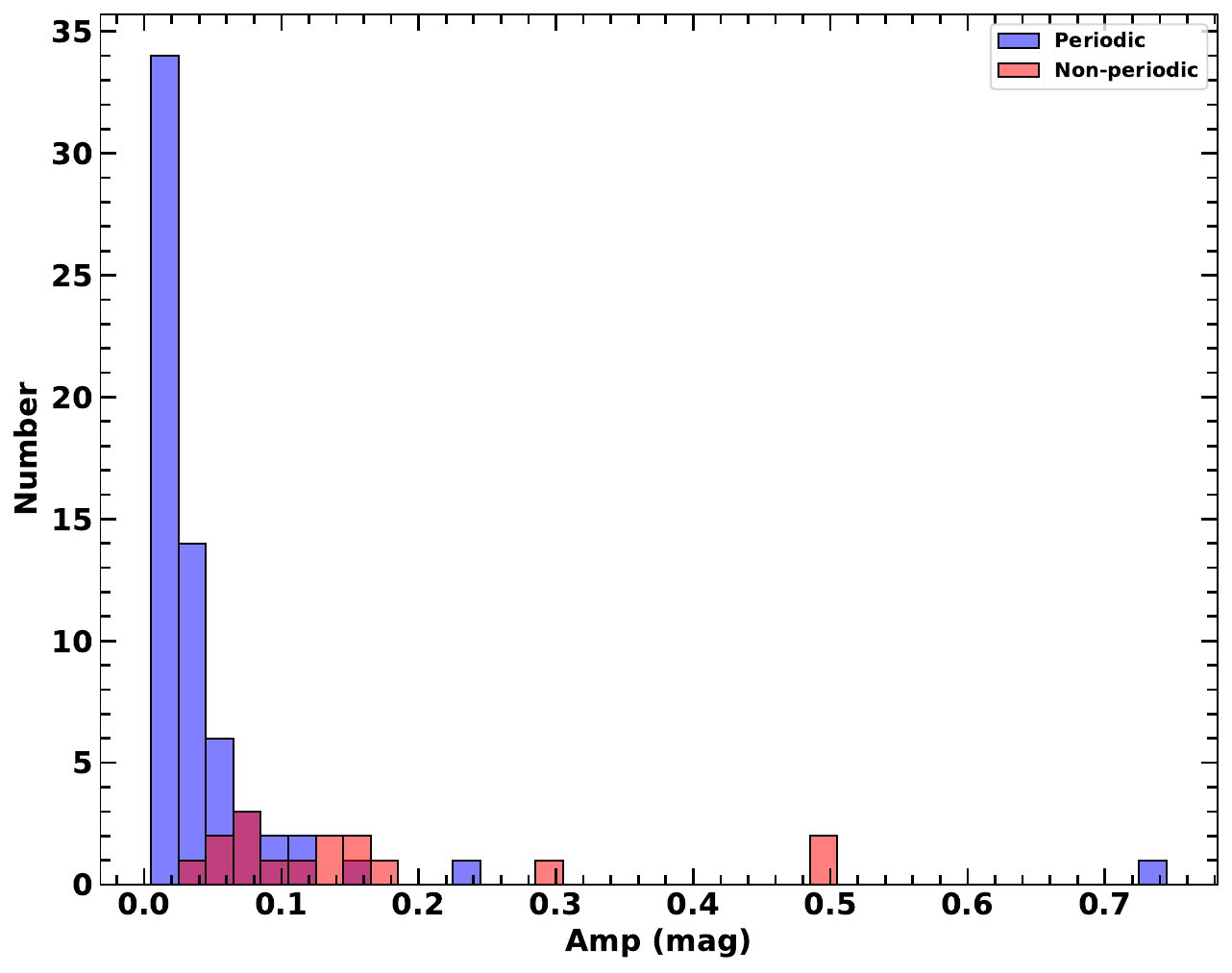}
    \includegraphics[width=0.45\linewidth]{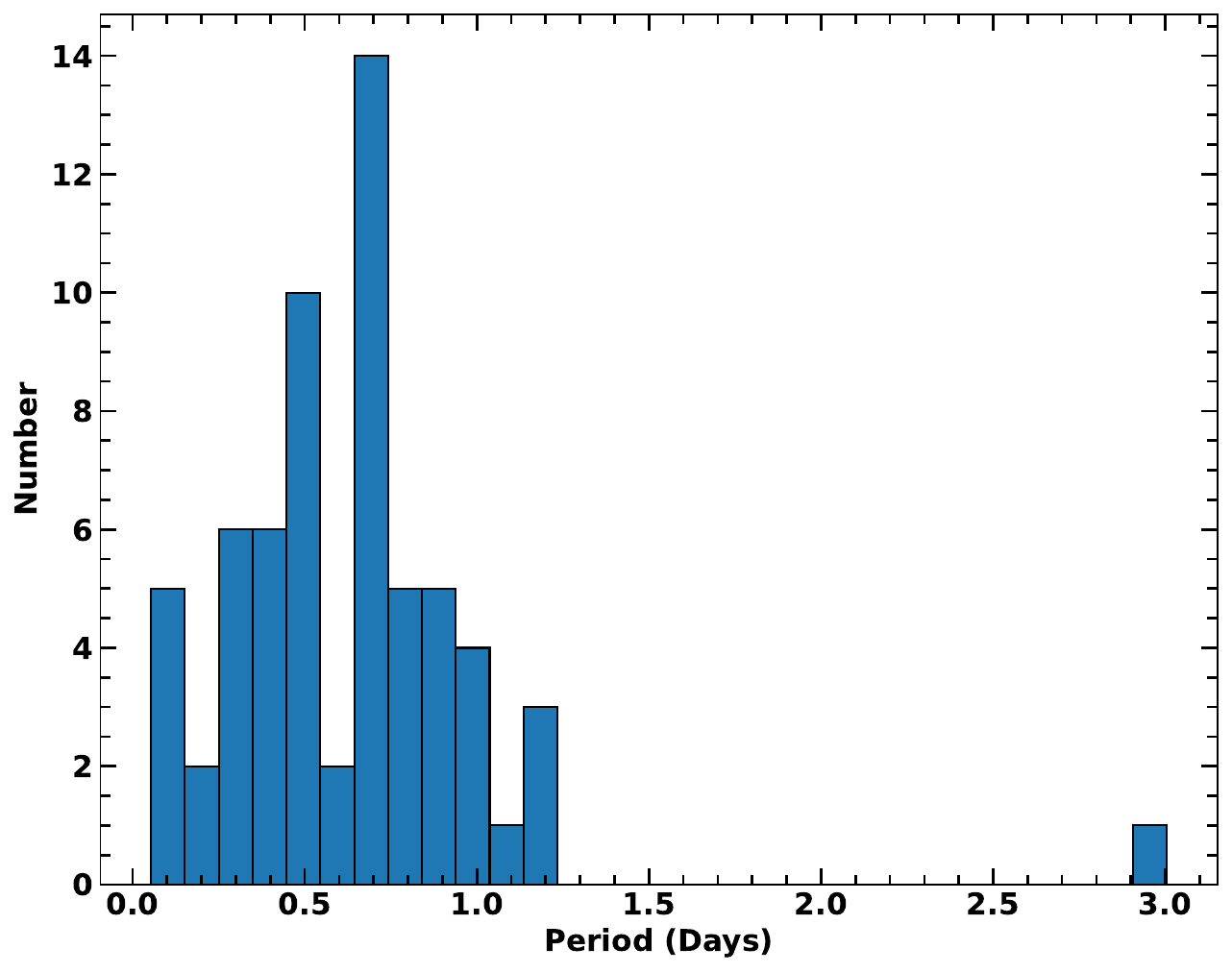}
    \caption{In the left panel, the histogram shows the amplitude distribution of variable stars. The right panel shows the period distribution of periodic stars.}
    \label{fig:histo}
\end{figure}

Figure~\ref {fig:histo} shows the distribution of amplitude and period of the variables identified in the present study. The amplitude and period of the periodic stars range from  $\sim$8 mmag to $\sim$700 mmag and $\sim$0.05 days to $\sim$3.00 days, respectively. At the same time, the amplitude of non-periodic stars ranges from $\sim$30 mmag to $\sim$500 mmag. The period distribution peaks around  0.7 days for all the periodic variables.
The amplitudes for non-periodic stars show a more or less flat distribution, whereas the periodic variables favor smaller amplitudes in their variability. 

The physical parameters, such as the age and mass of the cluster member variables, can be easily derived from their position in the CMD. However, it can't be done for other field variables as we don't know their exact extinction and distance. 
Thus, we utilized the \citet{Green_2019ApJ...887...93G}  map to determine the extinction values for these field variables. Subsequently, employing these extinction values along with the distances obtained from \citet{Bailer-Jones_2021AJ....161..147B}, we derived the absolute magnitude in the V band ($M_V$) for the field variables. By comparing the estimated $M_V$ values with the standard $M_V$ values in \citet{Pecaut_2013ApJS..208....9P}, we got the values of luminosity (log($L/L_\odot$)) and temperature (log($T_{eff}$)) for all the variables. In Figure~\ref{fig:hrd}, the Hertzsprung-Russell (HR) diagram (log($L/L_\odot$) vs. log($T_{eff}$)) is plotted for all the periodic variables in the FoV. 
In this HR diagram, the dotted blue line is the MS curve taken from  \citet{Pecaut_2013ApJS..208....9P}. 
The location of $\beta$ Cep, SPB, and $\delta$ Scuti stars in the HR diagram are represented by the dotted magenta line and solid black and blue lines, respectively \citep{Miglio_2007CoAst.151...48M, Balona_2011MNRAS.413.2403B}. The magenta dashed lines show the location of the $\gamma$ Dor stars \citep{Warner_2003ApJ...593.1049W}. The green dots represent the field periodic variables, whereas the red open circles represent the member periodic variables.
The distribution of the current sample of periodic variables in the HR diagram suggests that they have a large mass range (0.5-3.0 M$_\odot$); however, most are between $\sim$ 1-2 M$_\odot$.

\subsection{Classification of variables}
\label{sec:class_var}

 We have not detected any PMS signatures in the identified variables, as outlined in the study by \citet{Gutermuth_2009ApJS..184...18G, Koenig_2014ApJ...791..131K, Sharma_2016}. Thus, considering the identified variable as MS variables,  we tried to classify these variables into various known type classes. Our approach involved classifying these stars based on their period, variability amplitude, position in the HR diagram, and the shape of their LCs.

\begin{figure}[h!]
    \centering
    \includegraphics[width=8cm, height=8cm]{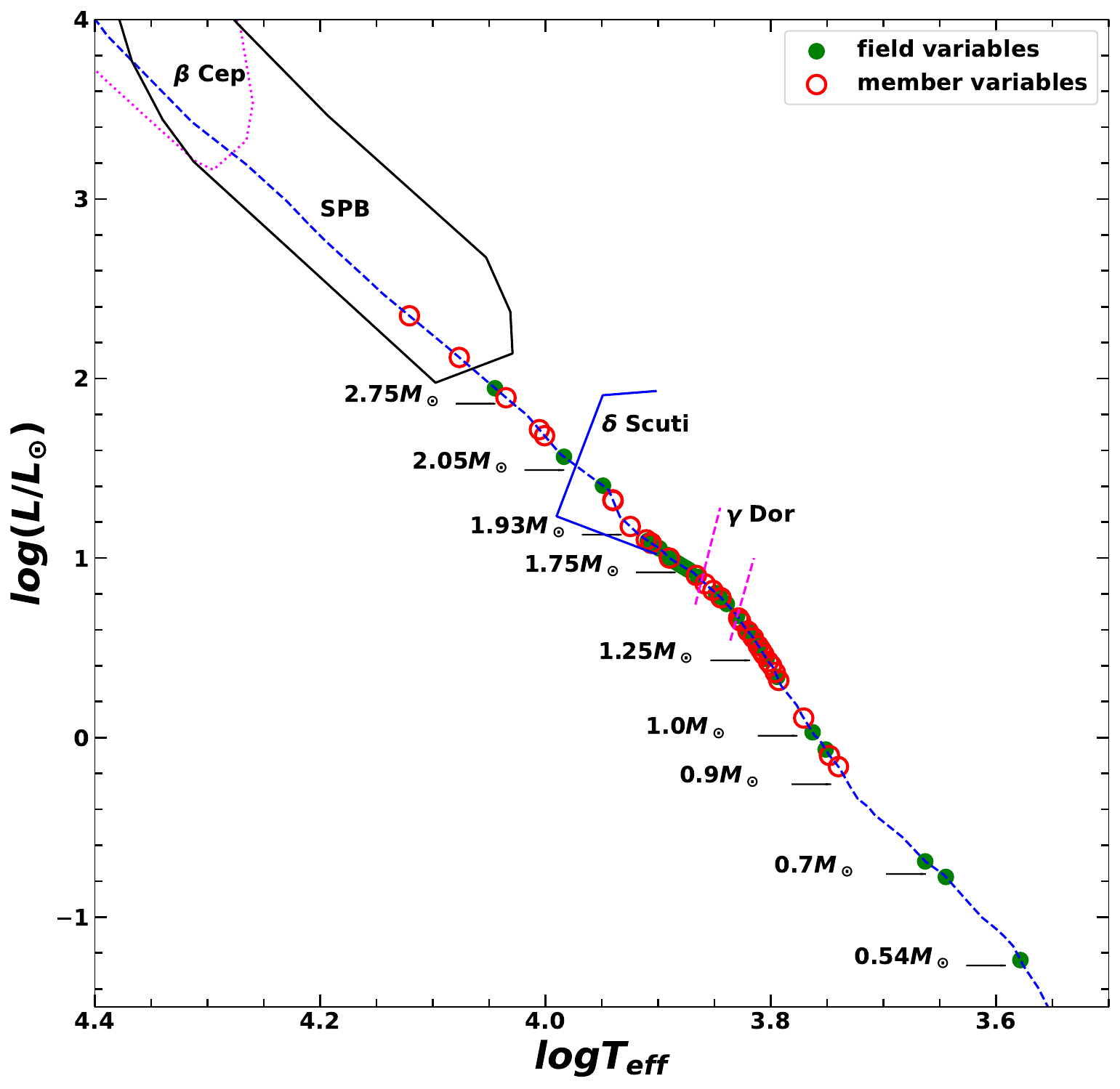}
    \caption{Hertzsprung-Russell (log($L/L_\odot$) vs. log($T_{eff}$)) diagram for periodic variables within the FoV. The dotted blue line is the MS curve from  \citet{Pecaut_2013ApJS..208....9P}. The green dots and red open circles represent the field and member periodic variables, respectively. The dotted magenta line shows the location of $\beta$ Cep-type MS variables, whereas the solid black line shows the location of SPB variables. The solid blue line shows the instability strip of $\delta$ Scuti stars, and the magenta dashed line shows the location of $\gamma$ Dor variables \citep{Warner_2003ApJ...593.1049W, Miglio_2007CoAst.151...48M, Balona_2011MNRAS.413.2403B}.}
    \label{fig:hrd}
\end{figure}

\begin{itemize}
    \item \textbf{Pulsating Variables :}
    $\beta$ Cep stars, which are pulsating MS variables, occupy a position above the upper MS on the HR diagram (dotted magenta line in Figure~\ref{fig:hrd}) and have early B spectral types \citep{Handler_2011}. They exhibit periods ranging from 0.1 to 0.6 days and amplitudes ranging from $\sim$10 mmag to $\sim$300 mmag. On the other hand, SPB stars are located just below the instability strip of $\beta$ Cep variables \citep{Waelkens_1991A&A...246..453W}. These stars display complex oscillations with multiple periods, attributed to non-radial g-mode pulsations driven by the $\kappa$ mechanism \citep{Gautschy_1993, Fedurco_2020A&A...633A.122F}. Typically, these variables have periods ranging from half a day to a few days in some instances \citep{Stankov_2005ApJS..158..193S}. $\delta$ Scuti stars are pulsating MS stars with spectral types A and F. They undergo radial and non-radial pulsations, with periods typically ranging from $\sim$ 0.008 days to 0.42 days \citep{2017A&A...597A..29S, 2023AAS...24117710M}. The amplitude of brightness variations is generally small, on the order of a few mmag to a few hundred mmag  \citep{Pietrukowicz_2020AcA....70..241P, Soszy_2021AcA....71..189S}. Another group of pulsation variables is known as $\gamma$ Dor, with periods ranging from several hours to a few days and amplitudes ranging from a few tens of mmag to hundreds of mmag. These stars are located below the instability strip of $\delta$ Scuti stars, and their instability strip overlaps with that of $\delta$ Scuti stars \citep{ Warner_2003ApJ...593.1049W,2018NewA...62...70I}.
    
    In our study, we identified two stars with periodic variations (V2 and V5) positioned within the SPB region on the HR diagram (refer to Figure~\ref{fig:hrd}). These stars have periods of 0.5058 $\pm$ 0.0008  days and 0.8678 $\pm$ 0.0015 days, respectively. Consequently, we classify these stars as SPB-type variable stars. The 19 periodic variables fall within the $\delta$ Scuti instability strip on the HR diagram. These stars exhibit periods that span from $\sim$0.09 days to $\sim$1.03 days, and their amplitudes range from $\sim$8 mmag to $\sim$90 mmag. Among them, 7 stars have periods that fall within the typical range of $\delta$ Scuti variables. Furthermore, the LCs of these stars are well in agreement with the LCs of  $\delta$ Scuti stars available in the literature \citep[see,][]{Pietrukowicz_2020AcA....70..241P}. Thus, these stars are marked as $\delta$ Scuti in the Table~\ref{tab:var}. The remaining 12 stars are marked as "RR Lyrae" stars as RR Lyrae stars have the same spectral type as $\delta$ Scuti but larger periods \citealp{Mullen_2023ApJ...945...83M, Gao_2025ApJS..276...57G}. The 9 identified periodic variables, with periods ranging from $\sim$0.21 days to $\sim$1.01 days and amplitudes ranging from $\sim$10 mmag to $\sim$43 mmag, are located within the $\gamma$ Dor instability strip in the HR diagram. These stars are classified as $\gamma$ Dor in Table~\ref{tab:var}.
    One periodic variable, V68, located just below the $\gamma$ Dor instability strip, has a period (0.0537 $\pm$ 0.0001 days; very short for rotating variable), amplitude ($\sim$15 mmag), and shape of the light curve, similar to that of a pulsating star. Considering the potential errors in the calculated luminosity and temperature values, we have marked this variable as "$\gamma$ Dor" in Table~\ref{tab:var}.

    \item\textbf{Non-pulsating Variables :}
   A group of five stars, V7, V10, V13, V22, and V55, are located within the gap between the SPB and $\delta$ Scuti instability regions. \citet{Mowlavi_2013A&A...554A.108M} found these types of stars in the open cluster NGC 3766, where they observed a significant population of newly identified variable stars occupying the region between SPB and $\delta$ Scuti stars, where standard stellar models do not predict any pulsation \citep{Balona_2011MNRAS.413.2403B, Mowlavi_2013A&A...554A.108M, Szewczuk_2017MNRAS.469...13S}. However, it has been observed that stars do exist in this gap and are classified as non-pulsating variables despite actually pulsating. This pulsation may be attributed to rapid rotation, which alters the internal conditions of the stars \citep{Balona_2011MNRAS.413.2403B}. The findings of \citet{Mowlavi_2013A&A...554A.108M} were further supported by \citet{Lata_2014MNRAS.442..273L, Lata_2019AJ....158...68L} in their study of the young clusters NGC 1893 and Stock 8. \citet{Mowlavi_2013A&A...554A.108M} reported that the periods of these variable stars range from 0.1 to 0.7 days, with amplitudes between 1 and 4 mmag. On the other hand, \citet{Lata_2014MNRAS.442..273L} identified a new class of variables in NGC 1893, with periods ranging from 0.17 to 0.58 days and amplitude variations between 7 and 19 mmag. In the case of Stock 8, \citet{Lata_2019AJ....158...68L}  identified similar variables, with amplitudes of up to 40 mmag and periods of up to 0.364 days. In this study, the newly identified class of variables have periods ranging from 0.09 to 0.74 days, with amplitudes ranging from 10 to 44 mmag. However, one newly identified class variable (V10) has a period of $\sim$ 3 days and an amplitude of 44 mmag. These variables are marked as "Non-pulsating" in Table~\ref{tab:var}.
   
    \item \textbf{Eclipsing Binary :}  The shape of the LC of one identified periodic variable, V62, suggests that this variable is the probable candidate for an eclipsing binary (EB). Based on the distinct primary and secondary minima observed in the LC of V62, it can be concluded that this star is a detached binary system. The period of the complete cycle for this EB is calculated as 1.2 days. The brightness variation during primary and secondary minima is calculated as approximately 800 mmag and 150 mmag, respectively. The phase-folded LC is given in the lower panel of Figure~\ref{fig:phase}. This star is listed in VizieR Online Data Catalog: KISOGP \citep{Ren_2021yCat..51610176R} as an Algol-type eclipsing binary with a period 1.2003209$\pm$0.0000029 days and with an eclipsing ratio 0.245. This star is also identified by Gaia DR3 \citep{Gaia_2022yCat.1358....0G} as EB with similar period and brightness variation in the G band.

    \item \textbf{Rotating Variables :} This section discusses those variable stars that are located below the $\gamma$ Dor instability strip in the HR diagram. The mass and period range (i.e., solar to subsolar and few hours to few days, respectively) of these variables are well in agreement with the BY Draconis (BY Dra) type variables. BY Dra stars are main-sequence stars belonging to the FGKM spectral types. They exhibit variable brightness caused by starspots, chromospheric activity, and other phenomena related to their magnetic fields \citep{Chahal_2022MNRAS.514.4932C}. These stars are quasi-periodic and change the shape of their LC over a month. The typical amplitude for the BY Dor variable is a few hundred mmag. The 26 identified periodic variables are classified as probable BY Dor in the present work. The period and amplitude of these 28 variables range from approximately 0.30 days to 1.20 days and 12 mmag to 225 mmag, respectively. Three of these variables, V48, V52, and V77, are identified by Gaia DR3 \citep{Gaia_2022yCat.1358....0G} as solar-like, RS Canum Venaticorum type and short-time variables, respectively. 

    \item \textbf{Non-periodic variables :} The sixteen identified variables are marked as non-periodic in the present study. Two are classified as cluster members, and fourteen are field populations (see Section~\ref{sec:identi_var}). The non-member variables show amplitudes ranging from $\sim$ 30 mmag to 500 mmag, and the member variables have amplitudes $\sim$ 60 mmag and 500 mmag. One of the member variables (V80) is classified as a long-period (with period $\sim$ 691.4704 days) variable by the Gaia variability pipeline \citep{Lebzelter_2023A&A...674A..15L}. However, our analysis categorized this variable as non-periodic due to insufficient data to determine its periodicity in the LC.
    The variables, V15, V31, and V57 are classified as solar-like variable in Gaia DR3 \citep{Gaia_2022yCat.1358....0G}. 
    The periods for V31 and V57 are mentioned as 1.67 days and 4.1 days, respectively.
\end{itemize}
One periodic variable star, V26, could not be classified due to the lack of distance information in \citealt{Bailer-Jones_2021AJ....161..147B}. Therefore, we have labeled it as 'Periodic/Field' in Table~\ref{tab:var}.
\section{Summary and conclusion}
\label{sec:summ}
We have used deep optical photometric data from the ARIES telescopes to study the physical properties of a poorly studied cluster, Be 65. The shape of this cluster is more or less circular, with an aspect ratio of 1.1. The center of the cluster is found to be at $\alpha_{2000} = 02^h  39^m 05^s $, $ \delta_{2000} = +60^\circ 24^{'} 30^{''} $ and the size and area of the cluster are estimated as $R_{cluster}$ = 1.6$^{'}$ (= 0.95 pc at a distance of 2 kpc) and $A_{cluster}$ = 8.3 $arcmin^2$, respectively. Using the PM data from Gaia DR3, we have identified members of this cluster. We have estimated this cluster's distance to 2.0 $\pm$ 0.1 kpc based on the parallax of member stars and isochrone fitting on optical CMD. The brightest member stars of the cluster are estimated to be 3.5 M$_\odot$ massive, and based on the turn-off point, the age of this cluster is estimated as $\sim$ 160 Myrs. We have found evidence of mass segregation in this cluster, and the comparison of the age of this cluster (160 Myr) with its dynamical age (7.5 Myr) suggests that the observed mass segregation can be due to the internal dynamical evolution of the cluster. The MF slope in this cluster's relatively higher-mass ($1.7 M_\odot \leq M \leq 3.5 M_{\odot}$) regime is estimated as -2.52, steeper than the Salpeter value of -1.35, and the MF shows a turn-off in its distribution at bit higher mass (at $\sim$1.7 M$_\odot$). The MF slope of +1.27 at the lowest mass bin ($ 0.8M_\odot \leq M \leq 1.7 M_\odot$) suggests fewer low-mass stars in this cluster. Thus, the overall MF distribution hints towards losing low-mass stars probably due to the dynamical evolution in the Be 65 cluster. The low value of the estimated photometric mass, i.e., $\sim$164 M$_\odot$, in comparison to its dynamical mass ($\sim$5581 M$_\odot$), confirms that this cluster has lost stellar mass and, thus, is under the process of disruption. A cluster can disrupt under the influence of both the internal and external dynamical evolution \citep[see also,][]{2019ApJ...877...12T}.
The stellar system can lose a substantial fraction of its stars in a time scale of $\tau_{\text{evap}} \sim 100 \times T_E$ due to internal dynamical evolution \citep{Shu_1982phyn.book.....S, Binney_1987gady.book.....B}. External disturbances, such as tidal forces from nearby giant molecular clouds or star clusters, passages through Galactic spiral arms or discs, or shear forces caused by Galactic differential rotation, can further speed up the cluster's disintegration. 
The Be 65 cluster, located within the Galactic disc in an environment of gas and dust (see Figure \ref{fig:rgb}), with an age of approximately 160 Myr, seems to have lost stars much earlier than its $\tau_{\text{evap}}$ ($\sim$0.75 Gyr) time scale, suggesting dominant roles of the external forces in its dynamical evolution.

Using a homogeneous decade-long baseline optical data, we also searched for the variables in the direction of the Be 65 cluster region. We have identified 64 periodic and 16 non-periodic stars, with periods ranging from 0.05 days to 3 days and amplitudes ranging from approximately 8 mmag to 742 mmag. Out of them, 35 are found to be members of the cluster Be 65, and the remaining ones are the field population.
Using the position in the HR diagram and the shape of the LCs, we have characterized the variability properties of the periodic stars. Of them, 31 are categorized as main-sequence pulsating variables of different types, such as SPB (2), $\delta$ Scuti (7), RR Lyrae (12), and  $\gamma$ Dor (10).
We have found 5 variables that do not follow the standard pulsation models and fall between SPB and $\delta$ Scuti instability regions. Their pulsation may be attributed to rapid rotation, which alters the internal conditions of the stars \citep{Balona_2011MNRAS.413.2403B}. We also found a detached binary system (V62) based on its distinct primary and secondary minima observed in the LC. We have identified 26 rotating variables with mass and period ranges (i.e., solar to subsolar and a few hours to a few days, respectively) similar to BY Dra-type variables. They exhibit variable brightness caused by starspots, chromospheric activity, and other phenomena related to their magnetic fields. 
The 16 identified variables are categorized as non-periodic, as we cannot phase-fold their light LCs. These variables show  30 mmag to 500 mmag amplitude range. A few of them (V15, V31, and V57) are classified as solar-like variables in Gaia DR3.

\section*{Conflict of Interest Statement}

The authors declare no conflict of interest.

\section*{Acknowledgments}
The observations reported in this paper were obtained by using the 1.3 m Devasthal Fast Optical Telescope (DFOT) and 1.04 m Sampurnand Telescope (ST), Nainital, India. This work made use of data from the Two Micron All Sky Survey (a joint project of the University of Massachusetts and the Infrared Processing and Analysis Center/California Institute of Technology, funded by the National Aeronautics and Space Administration and the National Science Foundation), and archival data obtained with the Spitzer Space Telescope and Wide Infrared Survey Explorer (operated by the Jet Propulsion Laboratory, California Institute of Technology, under contract with the NASA. This publication also made use of data from the European Space Agency (ESA) mission Gaia (https://www.cosmos.esa.int/gaia), processed by the Gaia Data Processing and Analysis Consortium (DPAC,https://www.cosmos.esa.int/web/gaia/dpac/consortium). D.K.O. acknowledges the support of the Department of Atomic Energy, Government of India, under project identification no. RTI 4002. A.V. acknowledges the financial support of DST-INSPIRE (No.: DST/INSPIRE Fellowship/2019/IF190550).

\appendix
\section{ Appendix}
\subsection{Cluster Membership}
\label{sec:memb}
Accurately identifying probable cluster members is crucial for reliably deriving cluster parameters from techniques like isochrone fitting of CMD and TCD. The presence of field star contamination introduces errors in cluster parameters. Using proper motions, kinematic studies of star clusters allow the identification of probable cluster members based on their distinct proper motion distribution relative to field stars in the vector point diagram (VPD; see Figure~\ref{fig:vpd_mem}). The Gaia DR3 data release provides precise parallax measurements for faint stars with limiting G $<$ 21 mag. Therefore, we have used PM data analog with G bands magnitude from Gaia DR3 to calculate the membership probability of stars located within $\sim$ 20$^{'}$ $\times$ 20$^{'}$  FoV around Be 65. We have plotted the PMs of stars within the cluster region on a VPD (panel 1 in Figure~\ref{fig:vpd_mem}). From VPD, it is clear that a prominent clump of stars with similar PMs is likely cluster members. The remaining scattered stars are likely field stars. Panel 2 in Figure~\ref{fig:vpd_mem} shows the CMD (G vs ($G_{BP}$ - $G_{RP}$)) for all stars in FoV, members, and for field stars in sub-panels 2(a), 2(b) and 2(c), respectively. As member stars show a well-defined MS, it confirms their higher probability of membership. Assuming the distance of 2.27 kpc (from WEBDA) for Be 65 and the radial velocity dispersion of 1 (km/sec) for the open cluster \citep{Girard_1989AJ}, we have calculated the expected PM dispersion ($\sigma_{c}$) for cluster members. Considering stars within a circle centered at $\mu_{RA}= -0.70 (mas/year)$, $\mu_{Dec}= -0.43 (mas/year)$ and radius of 0.3 (mas/year) on VPD, as probable members, we have derived frequency distribution $\sigma^{\nu}_{c}$ and $\sigma^{\nu}_{f}$ for member and field stars, respectively then we have calculated membership probability for stars in the field. The detailed procedure is described in \cite{Sharma_2020MNRAS.498.2309S}. The membership probability, associated errors in proper motion, and parallax values are plotted as a function of G magnitude in Figure~\ref{fig:vpd_mem} (panel 3). It is clear that the stars with a high membership probability ($P_{\mu} > 80\%$) exist down to faint magnitudes of G $\sim$20, and there is a clear separation between member and field stars at brighter magnitudes. The bottom sub-panel of panel 3 in Figure~\ref{fig:vpd_mem} shows the parallax of the stars as a function of G magnitude. With a few outliers, most stars with a high membership probability ($P_{\mu} > 80\%$) follow a tight distribution. The membership probability was estimated for 3517 stars in the cluster region, of which 540 stars were found to be cluster members ($P_{\mu} > 80\%$).

\bibliographystyle{Frontiers-Harvard} 
\bibliography{references}

\begin{table*}[h!]
    \centering
    \caption{Log of observations.}
    \begin{tabular}{c c c c c}
    \hline
    \hline
    Date of observation & Filter & Exp Time (in s) & No. of frames & Instrument \\
    \hline
    2005-10-28 & V & 900 & 3 & 2K$\times$2K ST \\
    2005-11-04 & U; B; V; R; I & 300; 180; 180; 60; 300 & 3; 3; 3; 3; 9 & 2K$\times$2K ST \\
    2014-11-17 & V & 60 & 460 & 2K$\times$2K DFOT \\
    2014-12-26 & V & 60 & 80 & 2K$\times$2K DFOT \\
    2016-10-21 & V & 60 & 3 & 2K$\times$2K DFOT \\
    2016-11-26 & V & 60 & 70 & 2K$\times$2K DFOT \\
    2016-11-27 & V & 60 & 240 & 2K$\times$2K DFOT \\
    2017-02-17 & V & 60 & 3 & 2K$\times$2K DFOT \\
    2017-02-18 & V & 60 & 3 & 2K$\times$2K DFOT \\
    2017-02-23 & V & 60 & 3 & 2K$\times$2K DFOT \\
    2017-02-24 & V & 60 & 3 & 2K$\times$2K DFOT \\
    2017-10-15 & V & 300 & 4 & 2K$\times$2K DFOT \\
    2017-10-16 & U; B; V; R; I & 480; 300; 10,300; 60; 10,300 & 3; 3; 3,3; 3; 3,3 & 2K$\times$2K DFOT \\
    2017-10-26 & V & 300 & 4 & 2K$\times$2K DFOT \\
    2017-11-12 & V & 300 & 5 & 2K$\times$2K DFOT \\
    2017-11-23 & V & 300 & 16 & 2K$\times$2K DFOT \\
    2017-11-24 & V & 300 & 4 & 2K$\times$2K DFOT \\
    2017-12-22 & V & 300 & 8 & 2K$\times$2K DFOT \\
    2017-12-23 & V & 300 & 2 & 2K$\times$2K DFOT \\
    2018-01-20 & V & 300 & 3 & 2K$\times$2K DFOT \\
    2018-01-21 & V & 300 & 4 & 2K$\times$2K DFOT \\
    2018-02-08 & V & 10,300 & 3,3 & 2K$\times$2K DFOT \\
    2018-12-05 & V & 20,300 & 4,4 & 2K$\times$2K DFOT \\
    2018-12-06 & V & 20,300 & 3,3 & 2K$\times$2K DFOT \\
    2019-10-26 & V & 20,300 & 2,2 & 2K$\times$2K DFOT \\
    2019-10-28 & V & 20,300 & 2,2 & 2K$\times$2K DFOT \\
    2022-10-17 & V & 180 & 122 & 2K$\times$2K DFOT \\
    2022-10-18 & V & 180 & 60 & 2K$\times$2K DFOT \\
    2022-11-04 & V & 180 & 20 & 2K$\times$2K DFOT \\
    \hline
    \end{tabular}
    \label{tab1}
\end{table*}

\footnotesize

\begin{longtable}[h!]{c c c c c c c c c c c c }
    \caption{Catalog of the variable stars}
    \label{tab:var} \\
    \hline
    ID & RA & Dec & V & eV & V-I & e(V-I) & Period & ePeriod & Amp & eAmp & Comment \\	
    & (deg) & (deg) & (mag) & (mag) & (mag) & (mag) & (days) & (days) & (mag) & (mag) & \\
    \hline
    \endfirsthead

    \multicolumn{12}{c}{\tablename\ \thetable{} -- Continued} \\
    \hline
    ID & RA & Dec & V & eV & V-I & e(V-I) & Period & ePeriod & Amp & eAmp & Comment \\	
    & (deg) & (deg) & (mag) & (mag) & (mag) & (mag) & (days) & (days) & (mag) & (mag) & \\
    \hline
    \endhead
    
    \hline
    \endfoot
    
    \hline
    \endlastfoot
     V1  & 39.720505  & 60.476321  & 13.259  & 0.004  & 1.994  & 0.009  &      ---  &      ---  & 0.050  & 0.010   &  Nonperiodic/Field  \\				
     V2  & 39.941536  & 60.419930  & 14.145  & 0.003  & 1.330  & 0.007  & 0.5058  & 0.0008  & 0.011  & 0.006   &    Periodic(SPB)/Member  \\			
     V3  & 39.946917  & 60.321945  & 14.205  & 0.003  & 2.378  & 0.006  &      ---  &      ---  & 0.100  & 0.015   &  Nonperiodic/Field  \\				
     V4  & 39.831255  & 60.392879  & 14.374  & 0.005  & 1.306  & 0.007  & 1.0058  & 0.0034  & 0.010  & 0.004   &     Periodic($\gamma$ Dor)/Field  \\		       
     V5  & 39.823895  & 60.386711  & 14.475  & 0.004  & 1.190  & 0.006  & 0.8678  & 0.0015  & 0.022  & 0.009   &    Periodic(SPB)/Member  \\			
     V6  & 39.935830  & 60.438067  & 14.712  & 0.004  & 1.535  & 0.006  & 0.9654  & 0.0030  & 0.010  & 0.003   &     Periodic($\gamma$ Dor)/Field  \\		       
     V7  & 39.628701  & 60.393836  & 14.810  & 0.005  & 1.430  & 0.007  & 0.0861  & 0.0001  & 0.032  & 0.017   &    Periodic(Non-pulsating)/Member  \\		       
     V8  & 39.863105  & 60.438568  & 14.928  & 0.004  & 1.295  & 0.006  & 0.4516  & 0.0004  & 0.016  & 0.006   &     Periodic(Rotating)/Field  \\		       			
    V9  & 39.692304  & 60.445317  & 15.109  & 0.005  & 1.264  & 0.006  & 0.8619  & 0.0009  & 0.028  & 0.011   &     Periodic(RR Lyrae)/Field  \\		       
    V10  & 39.736583  & 60.397980  & 15.113  & 0.021  & 1.197  & 0.026  & 3.0039  & 0.0094  & 0.044  & 0.010   &    Periodic(Non-pulsating)/Member  \\		       	     
    V11  & 39.922676  & 60.528708  & 15.143  & 0.035  & 1.205  & 0.045  & 0.6502  & 0.0011  & 0.013  & 0.006   &     Periodic(RR Lyrae)/Field  \\		       
    V12  & 39.844409  & 60.417058  & 15.152  & 0.004  & 1.565  & 0.006  &      ---  &      ---  & 0.070  & 0.010   &  Nonperiodic/Field  \\				
    V13  & 39.975849  & 60.444071  & 15.170  & 0.038  & 1.304  & 0.048  & 0.4793  & 0.0007  & 0.010  & 0.005   &    Periodic(Non-pulsating)/Member  \\		      	     
    V14  & 39.722888  & 60.418012  & 15.352  & 0.005  & 1.480  & 0.007  & 0.7267  & 0.0014  & 0.012  & 0.006   &     Periodic(Rotating)/Field  \\		       
    V15  & 39.739006  & 60.383476  & 15.539  & 0.005  & 1.808  & 0.007  &      ---  &      ---  & 0.170  & 0.020   &  Nonperiodic/Field  \\						
    V16  & 39.933530  & 60.372973  & 15.700  & 0.003  & 1.476  & 0.006  & 0.5757  & 0.0008  & 0.012  & 0.007   &     Periodic($\gamma$ Dor)/Field  \\		       
    V17  & 39.907136  & 60.544052  & 15.750  & 0.038  & 1.402  & 0.048  &      ---  &      ---  & 0.030  & 0.010   &  Nonperiodic/Field  \\				
    V18  & 39.946733  & 60.525744  & 15.846  & 0.048  & 2.222  & 0.062  & 0.9337  & 0.0012  & 0.011  & 0.005   &    Periodic(RR Lyrae)/Member  \\		       
    V19  & 39.792712  & 60.397589  & 15.858  & 0.005  & 1.359  & 0.007  & 0.9028  & 0.0021  & 0.014  & 0.004   &    Periodic(RR Lyrae)/Member  \\		       
    V20  & 39.747914  & 60.493748  & 15.887  & 0.005  & 1.475  & 0.009  & 0.7879  & 0.0014  & 0.013  & 0.006   &     Periodic(RR Lyrae)/Field  \\		       
    V21  & 39.641814  & 60.393076  & 15.913  & 0.003  & 1.410  & 0.006  &      ---  &      ---  & 0.080  & 0.015   &  Nonperiodic/Field  \\				
    V22  & 39.520161  & 60.475308  & 15.915  & 0.057  & 2.784  & 0.073  & 0.5056  & 0.0007  & 0.010  & 0.006   &     Periodic(Non-pulsating)/Field  \\		       
    V23  & 39.608363  & 60.409567  & 15.919  & 0.005  & 1.423  & 0.008  &      ---  &      ---  & 0.060  & 0.010   & Nonperiodic/Member  \\				
    V24  & 39.727561  & 60.305753  & 15.952  & 0.004  & 1.507  & 0.006  &      ---  &      ---  & 0.130  & 0.025   &  Nonperiodic/Field  \\				
    V25  & 39.712146  & 60.358012  & 15.998  & 0.005  & 2.290  & 0.009  & 0.6320  & 0.0007  & 0.015  & 0.008   &     Periodic(RR Lyrae)/Field  \\		       
    V26  & 39.932458  & 60.537526  & 16.035  & 0.038  & 1.470  & 0.049  & 0.9539  & 0.0010  & 0.016  & 0.008   &     Periodic/Field  \\				
    V27  & 39.699695  & 60.443812  & 16.121  & 0.004  & 1.445  & 0.008  & 1.0702  & 0.0018  & 0.020  & 0.009   &     Periodic(Rotating)/Field  \\		       
    V28  & 39.560095  & 60.363216  & 16.146  & 0.004  & 1.487  & 0.010  & 0.3223  & 0.0002  & 0.016  & 0.009   &     Periodic(Rotating)/Field  \\		       
    V29  & 39.605768  & 60.464853  & 16.170  & 0.003  & 1.165  & 0.008  & 0.3219  & 0.0002  & 0.013  & 0.009   &     Periodic($\delta$ Scuti)/Field  \\		       
    V30  & 39.604962  & 60.337468  & 16.198  & 0.012  & 1.874  & 0.017  & 0.3224  & 0.0001  & 0.013  & 0.008   &    Periodic($\delta$ Scuti)/Member  \\		       
    V31  & 39.514935  & 60.373143  & 16.252  & 0.048  & 2.212  & 0.062  &      ---  &      ---  & 0.160  & 0.020   &  Nonperiodic/Field  \\				
    V32  & 39.695983  & 60.367846  & 16.353  & 0.003  & 1.471  & 0.009  & 1.0286  & 0.0013  & 0.030  & 0.009   &    Periodic(RR Lyrae)/Member  \\		       
    V33  & 39.504900  & 60.339187  & 16.393  & 0.037  & 1.326  & 0.047  & 0.4958  & 0.0006  & 0.010  & 0.007   &    Periodic(RR Lyrae)/Member  \\		       
    V34  & 39.957744  & 60.412576  & 16.395  & 0.006  & 1.684  & 0.007  & 0.2154  & 0.0001  & 0.008  & 0.011   &    Periodic($\delta$ Scuti)/Member  \\		       
    V35  & 39.723520  & 60.463519  & 16.531  & 0.004  & 1.547  & 0.006  & 0.5201  & 0.0005  & 0.016  & 0.009   &     Periodic(Rotating)/Field  \\		       
    V36  & 39.776654  & 60.470334  & 16.592  & 0.006  & 1.515  & 0.012  & 0.3618  & 0.0001  & 0.029  & 0.015   &    Periodic($\delta$ Scuti)/Member  \\		       
    V37  & 39.940246  & 60.505081  & 16.594  & 0.005  & 1.526  & 0.034  & 0.7281  & 0.0008  & 0.015  & 0.009   &     Periodic(RR Lyrae)/Field  \\		       
    V38  & 39.815099  & 60.474716  & 16.716  & 0.004  & 1.266  & 0.008  & 0.7013  & 0.0006  & 0.025  & 0.011   &     Periodic(RR Lyrae)/Field  \\		       
    V39  & 39.954041  & 60.342575  & 16.741  & 0.004  & 2.595  & 0.006  &      ---  &      ---  & 0.125  & 0.015   &  Nonperiodic/Field  \\							
    V40  & 39.864716  & 60.393709  & 16.820  & 0.020  & 2.521  & 0.023  &      ---  &      ---  & 0.150  & 0.030   &  Nonperiodic/Field  \\				
    V41  & 39.791664  & 60.355647  & 16.833  & 0.008  & 1.937  & 0.010  & 0.1464  & 0.0001  & 0.016  & 0.012   &    Periodic($\delta$ Scuti)/Member  \\		       
    V42  & 39.737806  & 60.430669  & 16.902  & 0.007  & 1.577  & 0.011  & 0.0892  & 0.0001  & 0.024  & 0.013   &     Periodic($\delta$ Scuti)/Field  \\		       
    V43  & 39.766324  & 60.393774  & 16.942  & 0.009  & 1.445  & 0.013  & 0.7770  & 0.0010  & 0.021  & 0.012   &    Periodic($\gamma$ Dor)/Member  \\		       
    V44  & 39.653417  & 60.490077  & 17.047  & 0.004  & 1.689  & 0.008  & 0.2140  & 0.0001  & 0.014  & 0.014   &    Periodic($\gamma$ Dor)/Member  \\		       		
    V45  & 39.773688  & 60.285470  & 17.123  & 0.042  & 1.745  & 0.054  & 0.4554  & 0.0002  & 0.020  & 0.018   &     Periodic(Rotating)/Field  \\		       
    V46  & 39.705667  & 60.451790  & 17.149  & 0.006  & 1.498  & 0.011  & 0.7183  & 0.0005  & 0.031  & 0.014   &    Periodic($\gamma$ Dor)/Member  \\		       
    V47  & 39.694434  & 60.505336  & 17.150  & 0.021  & 1.482  & 0.067  & 0.7281  & 0.0005  & 0.039  & 0.013   &    Periodic($\gamma$ Dor)/Member  \\		       
    V48  & 39.567552  & 60.531938  & 17.172  & 0.046  & 1.897  & 0.058  & 0.6986  & 0.0006  & 0.027  & 0.010   &     Periodic(Rotating)/Field  \\		       
    V49  & 39.477496  & 60.317343  & 17.183  & 0.048  & 2.217  & 0.062  &      ---  &      ---  & 0.300  & 0.030   &  Nonperiodic/Field  \\				
    V50  & 39.795096  & 60.393275  & 17.254  & 0.004  & 2.637  & 0.008  & 0.4597  & 0.0003  & 0.017  & 0.012   &     Periodic($\gamma$ Dor)/Field  \\		       
    V51  & 39.905973  & 60.350667  & 17.278  & 0.004  & 1.808  & 0.007  & 0.4206  & 0.0002  & 0.026  & 0.013   &     Periodic(Rotating)/Field  \\		       
    V52  & 39.759961  & 60.481283  & 17.451  & 0.004  & 1.412  & 0.008  & 0.3857  & 0.0002  & 0.017  & 0.015   &    Periodic(Rotating)/Member  \\		       
    V53  & 39.936971  & 60.500374  & 17.466  & 0.015  & 1.602  & 0.027  & 0.8473  & 0.0004  & 0.043  & 0.015   &    Periodic($\gamma$ Dor)/Member  \\		       
    V54  & 39.856862  & 60.395369  & 17.495  & 0.024  & 1.798  & 0.042  & 0.6993  & 0.0003  & 0.049  & 0.021   &    Periodic(Rotating)/Member  \\		       
    V55  & 39.670907  & 60.418470  & 17.512  & 0.015  & 1.960  & 0.025  & 0.7428  & 0.0004  & 0.034  & 0.016   &     Periodic(Non-pulsating)/Field  \\		       
    V56  & 39.930378  & 60.469099  & 17.570  & 0.024  & 1.577  & 0.049  & 0.7281  & 0.0003  & 0.046  & 0.017   &     Periodic(Rotating)/Field  \\	
    V57  & 39.711173  & 60.486700  & 17.643  & 0.004  & 1.784  & 0.009  &      ---  &      ---  & 0.120  & 0.040   &  Nonperiodic/Field  \\				
    V58  & 39.607465  & 60.448024  & 17.646  & 0.005  & 1.726  & 0.007  & 0.4058  & 0.0002  & 0.017  & 0.017   &    Periodic(Rotating)/Member  \\		       
    V59  & 39.783238  & 60.469015  & 17.705  & 0.043  & 2.446  & 0.055  & 0.7281  & 0.0003  & 0.058  & 0.021   &     Periodic(Rotating)/Field  \\		       
    V60  & 39.675972  & 60.409314  & 17.743  & 0.005  & 2.112  & 0.011  & 1.1358  & 0.0008  & 0.059  & 0.016   &    Periodic(Rotating)/Member  \\		       
    V61  & 39.512832  & 60.549854  & 17.745  & 0.046  & 1.763  & 0.058  & 0.6543  & 0.0005  & 0.024  & 0.018   &    Periodic(Rotating)/Member  \\		       
    V62  & 39.957821  & 60.452358  & 17.781  & 0.008  & 1.872  & 0.015  & 1.2003 &  0.0001 &  0.743 &  0.149   &     Periodic(Algol-EB)/Field  \\		       
    V63  & 39.942505  & 60.538305  & 17.850  & 0.050  & 1.653  & 0.060  & 0.7281  & 0.0002  & 0.073  & 0.025   &    Periodic(Rotating)/Member  \\		       
    V64  & 39.525307  & 60.286714  & 17.918  & 0.050  & 2.313  & 0.064  & 0.5115  & 0.0001  & 0.111  & 0.023   &    Periodic(Rotating)/Member  \\		       
    V65  & 39.827212  & 60.541931  & 17.974  & 0.053  & 1.833  & 0.066  & 0.5320  & 0.0002  & 0.053  & 0.025   &     Periodic(RR Lyrae)/Field  \\		       
    V66  & 39.721142  & 60.515226  & 17.976  & 0.045  & 1.761  & 0.056  & 0.7464  & 0.0004  & 0.045  & 0.027   &    Periodic(Rotating)/Member  \\		       
    V67  & 39.748481  & 60.365923  & 18.071  & 0.018  & 2.124  & 0.032  & 0.3543  & 0.0001  & 0.120  & 0.078   &    Periodic(Rotating)/Member  \\		       
    V68  & 39.832926  & 60.508947  & 18.131  & 0.011  & 1.876  & 0.014  & 0.0537  & 0.0001  & 0.015  & 0.018   &    Periodic($\gamma$ Dor)/Member  \\		       
    V69  & 39.625053  & 60.437254  & 18.170  & 0.008  & 2.165  & 0.015  & 0.4238  & 0.0002  & 0.025  & 0.022   &     Periodic(Rotating)/Field  \\		       
    V70  & 39.945166  & 60.533288  & 18.230  & 0.047  & 1.790  & 0.058  & 0.7281  & 0.0004  & 0.033  & 0.016   &    Periodic(Rotating)/Member  \\		       
    V71  & 39.500473  & 60.410428  & 18.345  & 0.080  & 1.424  & 0.090  & 0.3212  & 0.0001  & 0.077  & 0.057   &    Periodic(Rotating)/Member  \\		       
    V72  & 39.743180  & 60.374723  & 18.675  & 0.007  & 2.194  & 0.018  &      ---  &      ---  & 0.500  & 0.080   &  Nonperiodic/Field  \\				
    V73  & 39.794485  & 60.467601  & 18.679  & 0.029  & 2.564  & 0.035  & 0.6908  & 0.0002  & 0.089  & 0.049   &     Periodic(RR Lyrae)/Field  \\		       
    V74  & 39.684505  & 60.440382  & 18.903  & 0.019  & 2.207  & 0.045  & 0.7885  & 0.0001  & 0.225  & 0.070   &    Periodic(Rotating)/Member  \\		       
    V75  & 39.805476  & 60.421299  & 19.468  & 0.013  & 1.950  & 0.022  & 0.3016  & 0.0001  & 0.068  & 0.048   &    Periodic(Rotating)/Member  \\		       
    V76  & 39.865204  & 60.433041  & 19.642  & 0.010  & 2.412  & 0.013  & 1.2054  & 0.0005  & 0.161  & 0.040   &    Periodic(Rotating)/Member  \\		       
    V77  & 39.487082  & 60.465377  & 19.779  & 0.066  & 2.720  & 0.080  & 0.2951  & 0.0001  & 0.100  & 0.062   &     Periodic(Rotating)/Field  \\		       
    V78 & 39.620119  & 60.289778  &   13.437  &   0.015  &   ---  &   ---  &      ---  &      ---  & 0.065  & 0.020   &  Nonperiodic/Field  \\			
    V79  & 39.613911  & 60.506919  &    16.448  &   0.015  &   ---  &   ---  & 0.1369  & 0.0001  & 0.011  & 0.012   &     Periodic($\delta$ Scuti)/Field  \\	       	
    V80  & 39.519403  & 60.430814  &    16.187  &   0.020  &   ---  &   ---  &      ---  &      ---  & 0.500  & 0.030   & Nonperiodic/Member  \\ 
    \hline
\end{longtable}
\normalsize


\renewcommand\thefigure{A\arabic{figure}}
\setcounter{figure}{0}
\begin{figure}[h!]
        \includegraphics[width=7.8cm, height=10cm]{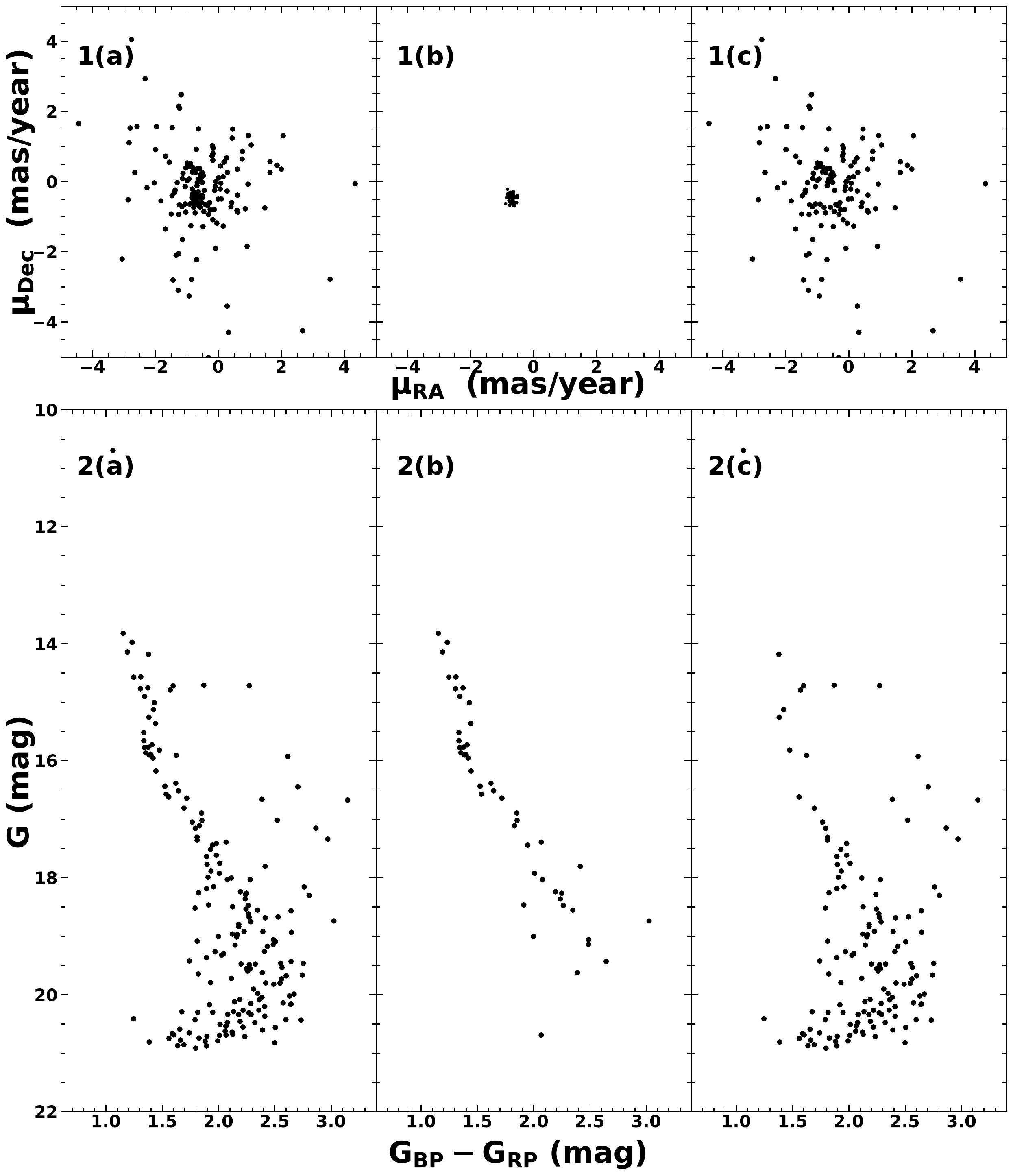}
        \includegraphics[width=9.8cm, height=10cm]{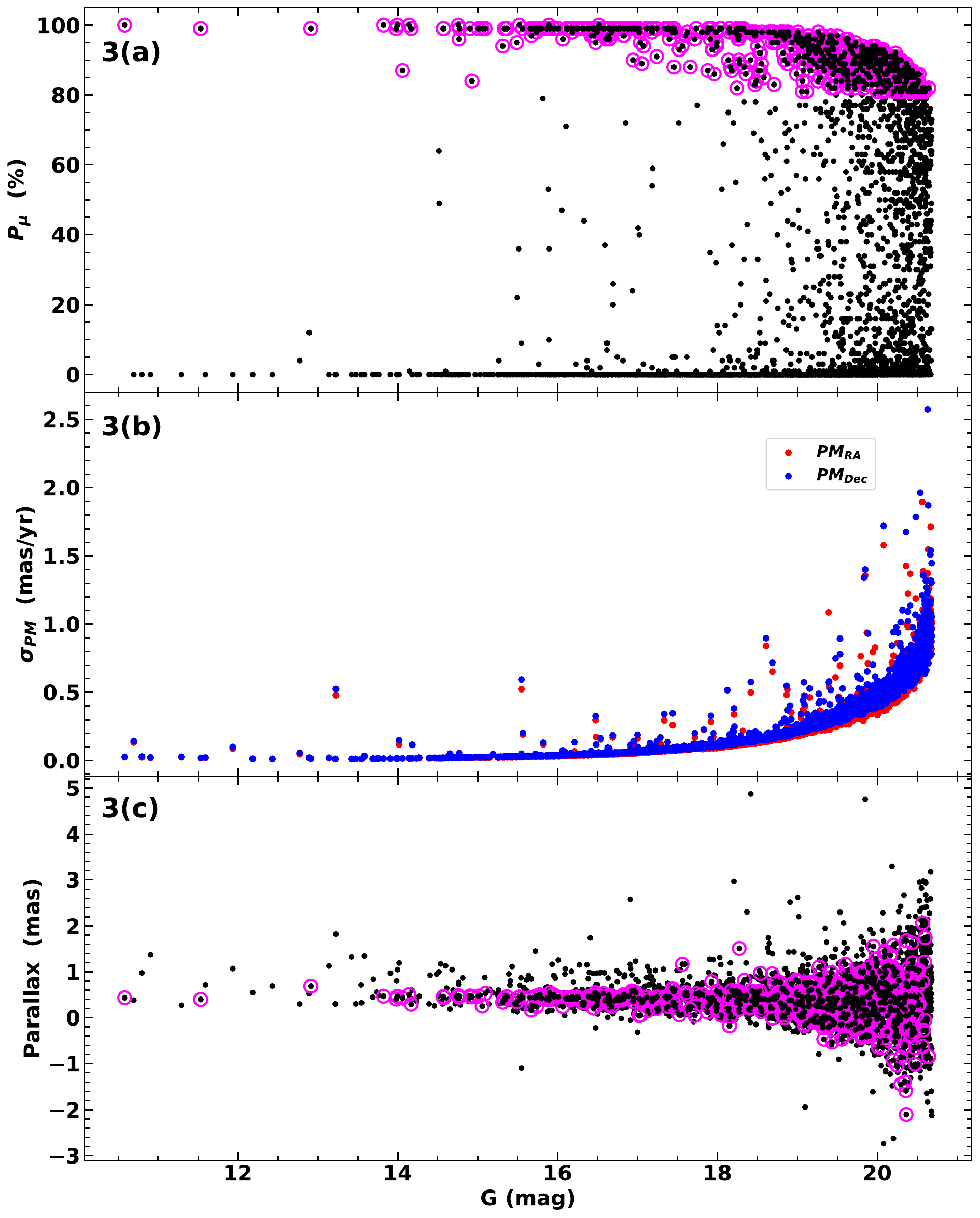}
    \caption{Left panels ( \textbf{1} and \textbf{2}): PM VPD (panel \textbf{1}) and Gaia DR3 G vs ($G_{BP}$ - $G_{RP}$) CMDs (panel \textbf{2}) for stars located in the Be 65 cluster region. The left sub-panels (\textbf{1(a)} and \textbf{2(a)}) show all stars in the cluster, while the middle (\textbf{1(b)} and \textbf{2(b)}) and the right sub-panels (\textbf{1(c)} and \textbf{2(c)}) show the probable cluster members and field stars, respectively. Right panels \textbf{(3)}: Membership probability P$\mu$ \textbf{(3(a))}, PM errors ($\sigma_{PM}$) \textbf{(3(b))}, and parallax \textbf{(3(c))} of stars as a function of G magnitude for stars in the Be 65 cluster region. 540 stars with $P_{\mu} > 80\%$ are considered members of the Be 65 cluster and are shown by circles with magenta rings.}
    \label{fig:vpd_mem}
\end{figure}

\end{document}